\documentclass[a4paper,11pt]{article}
\pdfoutput=1 % if your are submitting a pdflatex (i.e. if you have
\usepackage{jinstpub} % for details on the use of the package, please
\usepackage{multirow}
\usepackage{lineno}
\usepackage{chemfig}
\usepackage{array}
\usepackage{tikz}
\usepackage{graphicx}
\newcommand{\ovbb}{$0\nu\beta\beta$}
\newcommand{\tvbb}{$2\nu\beta\beta$}

\newcommand\subsubsubsection[1]{\paragraph{#1}}
\newlength{\capwidth}
\newlength{\chemwidth}
\newcommand\chemnamefullcap[2]
{%
	\settowidth{\capwidth}{#2}%
	\settowidth{\chemwidth}{#1}%
	\chemname*%
		{%
			\ifdim\capwidth > \chemwidth%
				\makebox[\capwidth][c]{#1}%
			\else%
				#1%
			\fi%
		}%
		{#2}%
}

\newcommand\chemnamewide[3]
{%
	\settowidth{\capwidth}{#2}%
	\settowidth{\chemwidth}{#1}%
	\chemname*%
		{%
			\ifdim\capwidth > \chemwidth%
				\makebox[1.2\capwidth][c]{#1}%
			\else%
				#1%
			\fi%
		}%
		{#2\\#3}%
}

\title{Gaseous forms of $^{76}$Ge, $^{82}$Se, $^{96}$Zr, $^{100}$Mo, $^{124}$Sn, and $^{130}$Te: new avenues to future \ovbb~time projection chambers}

\author[a]{A. Avasthi,}
\author[a]{B. Monreal,\note{Corresponding author.}}
\author[a]{I. Moya}

\affiliation[a]{Department of Physics, Case Western Reserve University, Cleveland OH}

\emailAdd{benjamin.monreal@case.edu}

\abstract{Searches for neutrinoless double beta decay are growing larger, with tonne-scale targets in several nuclides still far from exhausting the discovery space.  What's beyond ton scale?   Time projection chambers (TPCs) are one option for building large (100~T or kiloton-scale) instruments, but filling them with the familiar $^{136}$Xe for a $0\nu\beta\beta$ search is a problem: xenon is a scarce element whose atmospheric-extraction supply chain is small and hard to grow.  If future $0\nu\beta\beta$ searches wish to exploit TPCs' known hardware scalability, we need to fill them with non-xenon target materials.  Of particular value would be a TPC that can drift electrons, rather than ions, letting us use mature readout schemes which require gas gain.  In this paper, we identify a set of previously-unappreciated, affordable gases which are likely to be electropositive, allowing electron drift and gain in gas-phase TPCs sensitive to $0\nu\beta\beta$ with the help of track-topology background rejection.  We identify candidate $^{76}$Ge, $^{82}$Se, $^{96}$Zr, $^{100}$Mo, $^{124}$Sn, and $^{130}$Te compounds suitable for gas-phase electron-drift TPCs; some may be suitable for liquid-phase TPCs as well.  Using a figure-of-merit that emphasizes the need for track topology for background rejection, we argue that 100~T and kiloton-scale gas TPCs are realistic without unprecedented underground infrastructure. }
\begin{document}

\maketitle

\section{Introduction}

Neutrinoless double beta decay is one of the most interesting rare-event searches in modern nuclear physics.  Any discovery of a nonzero $0\nu\beta\beta$ rate, in any nuclide, would conclusively show that neutrinos are massive Majorana particles and that lepton number is not conserved.  Precise measurement of any half-life would be interpretable as a measurement of the neutrino mass scale.  Cross-checks between multiple nuclides, or measurement of the $\beta$-$\beta$ energy/angle distributions, could pin down this interpretation or add sensitivity to new physics.  However, experiments so far, despite multi-100-kg isotopic targets and near-zero backgrounds, have shown only that all $0\nu\beta\beta$ half-lives are extremely long.

While semiconductor, bolometer, scintillator, and tracking detectors have also accomplished or planned ton-scale \ovbb~searches, this paper's focus is on time projection chambers (TPCs).  TPCs have historically scaled up well: a 1000~kg experiment is not 10$\times$ more difficult than a 100~kg experiment.  Most of today's \ovbb~TPCs are searching for $^{136}$Xe decay.  Xenon is a familiar, clean substance with bright primary and secondary scintillation, useful in gas, two-phase, and single-phase liquid configurations\cite{albertSensitivityDiscoveryPotential2018,aalbersNeutrinolessDoubleBeta2025,abdusalamabdukerimPandaXxTADeepUnderground2025}.  Unfortunately, when considering beyond-10-ton-scale experiments like Origin-X, xenon's cost and supply limitations\cite{avasthiKilotonscaleXenonDetectors2021,ankerReportWorkshopXenon2024} may prohibit progress.  The only global source of xenon is the atmosphere, where it is a rare trace gas.  Xenon is currently available commercially only as a byproduct of cryogenic air separation plants that produce oxygen for the steel industry, so the supply is constrained by the scale of that operation; it would be extremely expensive to build or operate new xenon separation capacity.  Separation by adsorption on metal-organic frameworks\cite{kimHighYieldLargescale2024} is promising but probably not inexpensive at scale.  

In this paper, we ask whether future TPCs could adopt non-xenon fluids whose supply chains would allow easy access to 10 tonne, 100 tonne and kilotonne scales.  The kiloton scale is motivated by the underlying neutrino physics; the lightest possible Majorana neutrinos are {\em almost} guaranteed to have an effective neutrino mass $m_{\beta\beta} > 1$~meV, which is {\em almost} guaranteed to produce finite count rates in kiloton-scale detectors\cite{avasthiKilotonscaleXenonDetectors2021}.

The untested candidate fluids described here may have any number of detector disadvantages with respect to xenon, including modest to poor bulk radiopurity and lack of scintillation.  We hope that their disadvantages can be overcome via some or all of the following approaches:

\begin{enumerate}
\item Using gases rather than liquids.   This is a stringent constraint because the list of near-room-temperature gases containing \ovbb~target atoms is surprisingly small.  We anticipate two benefits to gas TPCs over liquids: 
  \begin{itemize}
\item Low Fano factors would allow us to improve energy resolution\cite{collaborationDemonstrationSubPercentEnergy2026}, narrowing the \ovbb~energy ROI. 
\item Gas TPCs allow high-spatial-resolution modes (as pioneered by NEXT \cite{martin-alboSensitivityNEXT100Neutrinoless2016}, Panda-X III\cite{chenPandaXIIISearchingNeutrinoless2017}, or AXEL\cite{hikidaAXELHighpressureXenon2025}) in which track topology provides signal/background discrimination\cite{simonBoostingBackgroundSuppression2021}.  In particular, the choice of pressure\cite{mistryOptimalOperatingParameters2025} provides an adjustable parameter that can be tuned towards higher discrimination power.
  \end{itemize}
\item Using isotopes whose \ovbb~endpoint is high enough to be above $^{232}$Th-chain contamination.
  \item Scaling up to even larger target masses.
\end{enumerate}

Before tackling the engineering and infrastructure needs of a giant non-xenon gas TPC, we have to ask what gases are available.  In this paper, we have done a careful and moderately comprehensive survey of chemical compounds (a) containing double beta decay target elements and (b) with melting and/or boiling below around 100~C.   Although there are many complexities to the question of which fluids will ultimately be suitable for large TPCs, these physical properies are rare enough that they provide a useful starting place.  After identifying these gases, we discuss some of the design criteria for giant gas TPCs and use a novel figure-of-merit to compare the candidate gases.

\section{Identifying gaseous compounds of \ovbb~target elements}

\subsection{Electropositive vs. electronegative gases}

There is a key experimental distinction between electropositive and electronegative compounds.  In an electronegative gas, primary ionization electrons form bound anion states by attaching to neutral gas molecules.  The anions might form via attachment (e$^-$ + SF$_6 \rightarrow$ SF$_6^-$, for example) or via dissociative attachment (DEA) (e$^-$ + SF$_6 \rightarrow$ SF$_5^-$ + F, for example), or both.  In either case, these anions drift slowly and must be detected without the benefit of gas amplification\cite{nygrenNeutrinolessDoubleBeta2018}.  Un-amplified ion current readouts suitable for high energy resolution, like TOPMETAL\cite{nvdex}, are in their infancy.   By contrast, in an electropositive gas, where the charge carriers are highly mobile free electrons, electrons can be drifted quickly.   If free electrons can be brought into high-electric-field regions, either near wires or in micropattern structures, then they can be multiplied through Townsend avalanches.  Gas gain (and, relatedly, electroluminescent) readout instrumentation involves mature technologies which are comparatively easy to build at the scales required here.  Electropositive gases are the focus of this paper.

Previous work on large non-noble TPCs\cite{mcconkeyLiquidScintillatorTime2013,mcconkeyLiquidScintillatorTime2014} focused largely on the challenges of the liquid phase.  Electron mobility in liquids is complicated because electron wavefunctions tend to extend over multiple molecules; drift can be slowed by collective or multibody effects.  The general rule is that polar liquids drift electrons extremely slowly\cite{jay-gerinMigrationExcessElectrons1991}, nonpolar liquids drift them faster,  with more-nearly-spherical nonpolar molecules\cite{schmidtElectronMobilityNonpolar1977,holroydElectronsNonpolarLiquids2003} being particularly fast.  Ultra-slow drift velocities are undesirable for a high-energy-resolution TPC because they allow time for recombination of the primary ionization cloud.  In gas TPCs, electron transport is much simpler; roughly speaking {\em all} electropositive gas molecules should allow drift with usable mobility.  

Not all electropositive gas species can support gain avalanches; electron-molecule collisions in or near the gain region are of much higher energy than those in the drift region, and most non-noble-gas molecules will undergo dissociation and attachment in addition to avalanche-supporting ionization.  (This phenomenon is familiar in collider experiments where gain avalanches can convert hydrocarbon drift gases into solid organic deposits on anode wires.)  We defer the investigation of these gases' gain behavior, and the possible engineering interventions to work around them, for future work.

\subsubsection{Electron affinities via density functional theory}

From a chemical perspective, electropositive gases are those that avoid electron attachment over the relevant drift distances.  In lieu of asking the complicated question of which gases have low attachment {\em cross sections}, we are asking the simpler and more conservative question of whether attachment is energetically favorable in near-thermal collisions.  Drifting electrons will have thermal or near-thermal energy distributions, so attachment only occurs if the process e$^-$ + M $\rightarrow$ M$^-$ is exothermic, i.e. if the binding energy of the neutral molecule E(M) is smaller than the binding energy of the anion E(M$^{-}$).  This quantity E(M) - E(M$^-$) is called the {\em electron affinity} (EA).  

For some common molecules, electron affinity can be found tabulated in plasma physics and chemical databases.  For less common molecules, we can calculate EAs using density functional theory (DFT).  In such calculations, we can choose between the ``adiabatic electron affinity'', comparing the neutral ground state to the anion ground state, or the ``vertical electron affinity'' (VEA), which calculates the anion energy without perturbing the positions of the nuclei from the neutral state.   In this paper, we have used the ORCA package\cite{neesefORCAProgramSystem2012} to model molecules with the B3LYP functional and the def2-TZZVP basis sets\cite{weigendBalancedBasisSets2005,rappoportPropertyoptimizedGaussianBasis2010,weigendAccurateCoulombfittingBasis2006}.   For small molecules, we are generally reporting adiabatic electron affinities.  For larger molecules, we settle for VEAs.  When possible, we have cross-checked our DFT calculations against previously published work; a table of examples are tabulated in the appendix.

In simple cases, DFT can predict whether DEA is energetically allowed by calculating the total binding energy of various possible combinations of fragments.  

\subsection{Candidate electropositive gases}

In this work, we have compiled a novel list of candidate electropositive gases.   Although little or nothing is known about the practicalities of any of these gases as drift media, we hope that the results shown below will motivate further study.  Without exception, the candidates are {\em toxic and flammable}; we hope to motivate research on engineering safeguards and underground infrastructure that would allow their safe use.

  \begin{table}
    \begin{center}{\bf Electron-drift \ovbb~TPC gas candidates}
    \begin{tabular}{c p{2cm} p{4cm} r r r}
      Nuclide & \multicolumn{2}{c}{Compound} & m.p.& b.p.& EA\\
              &   &                          & (C)   & (C)  & eV \\
    \hline
    \multirow{7}{*}{$^{82}$Se} & H$_2$Se & Hydrogen selenide & -66 & -41  &  -0.93 \\
    & CSeO & Carbonyl selenide\textsuperscript{$\star$} & -122 & -23 & -0.28 \\
    &     CH$_3$SeH & Methaneselenol & -54 & 12 & -0.67 \\
    &     H$_4$C$_4$Se & Selenophene & -38 & 110 &  -0.63\\
    &     H$_8$C$_4$Se & Selenolane & ?? & 135 &  -0.76\\
    &     C$_6$H$_5$SeH & Benzene selenol & ?? & 183 & -0.51\rlap{$^V$} \\
    &     C$_4$H$_8$Se$_2$ & 1,4-diselenane\textsuperscript{$\star\Updownarrow$} & 113 & 237 & -0.58 \\
    &    C$_6$H$_6$Se$_2$ & Benzene 1,4-diselenol\textsuperscript{$\star\Updownarrow$} & $>$97 & ? & -0.36 \\    
    \hline
     \multirow{2}{*}{$^{130}$Te}  &  H$_4$C$_4$Te & Tellurophene & -36  & $\sim$148 & -0.26 \\
                                 &  H$_8$C$_4$Te & Tellurolane & ? & $\sim$166 & -0.34 \\
    \hline
 \multirow{2}{*}{$^{76}$Ge}  &  GeH$_4$ & Germane\textsuperscript{$\Updownarrow$} & -165 & -88 & -1.27\\
 & Ge(CH$_3$)$_4$ & Tetramethyl~germanium\textsuperscript{$\Updownarrow$} & -88 & 43  & -1.02\rlap{$^V$} \\
 \hline
 \multirow{2}{*}{$^{124}$Sn}                           &  SnH$_4$ & Stannane\textsuperscript{$\star$$\Updownarrow$} & -146  & -52 & -0.90 \\
 & Sn(CH$_3$)$_4$ & Tetramethyl tin\textsuperscript{$\Updownarrow$} & -54 & 75  & -0.67\rlap{$^V$} \\
 \hline
 \multirow{3}{*}{$^{96}$Zr} & ((CH$_3$)$_2$N)$_4$Zr & Tetrakis(dimethylamido) zirconium(IV)\textsuperscript{$\star$$\Updownarrow$} & 60 & 89? & -0.75\rlap{$^V$}\\
 & ((CH$_3$)$_2$N)$_4$Zr & Zirconium(IV) tert-butoxide \textsuperscript{$\star$$\Updownarrow$}& 3 & 250? & -0.54\rlap{$^V$} \\
 & Zr(OC$_2$H$_5$)$_4$ & Zirconium(IV)ethoxide\textsuperscript{$\star$$\Updownarrow$} &  172 & $>$235 & -0.79\rlap{$^V$}\\
 \hline
 \multirow{1}{*}{$^{100}$Mo} & (C$_8$H$_{10}$)$_2$Mo & Bis(ethylbenzene) Mo\textsuperscript{$\star$$\Updownarrow$} & $<$23 & $>$170 & -0.76\rlap{$^V$}\\
 
    \end{tabular}
    \end{center}
    \caption{Summary of candidate electropositive drift gases identified for \ovbb~TPCs, with our DFT-calculated electron affinities, for which we suspect uncertainties of 0.3--0.5~eV.  Superscript-V$^V$ indicates a vertical electron affinity; all others are adiabatic.  Starred\textsuperscript{$\star$} compounds have unclear thermal stability.  Superscript-arrow\textsuperscript{$\Updownarrow$} compounds are nonpolar and plausibly interesting in the liquid state.}
  \end{table}\label{tab_electropositive}

The following subsections are ordered from most-plausible to least-plausible.   We start with small molecules but move towards larger ones when necessary, noting that Georges Charpak once operated gas detectors with trace ethyl ferrocene\cite{charpak1989} (C$_{12}$H$_{14}$Fe). We are not at this time looking for scintillating compounds. 

\subsubsection{Selenium}
$^{82}$Se ($Q_{\beta\beta}=2.996$ MeV, $f$=9.20\%) is the only non-xenon \ovbb~target previously considered for TPCs\cite{nvdex,nygrenNeutrinolessDoubleBeta2018} (as $^{82}$SeF$_6$).  Selenium itself is inexpensive and abundant; something like 3,500 T/y is extracted from copper ores with a value around USD20/kg.   As a chalcogen, selenium shares much of the chemistry of sulfur (and to some extent oxygen) and for that reason is readily found in liquid-phase or gas-phase forms.  For gas TPC purposes we will consider polar compounds with boiling points up to 135~C, and for liquid TPC purposes we will consider compounds with melting points up to 113~C.

Organoselenium compounds have a reputation, beyond that of sulfur, for foul odors; the discovery of CSe$_2$ released a stench that reportedly forced a small village to evacuate\cite{grimmUberDarstellungUnd1936}.   
    
\subsubsubsection{H$_2$Se}

\begin{center}
\chemname{\chemfig{Se([:-30]-H)([:200]-H)}}{Hydrogen selenide\\H$_2$Se\\m.p.\ -65.73, b.p.\ -41.25 C}
\end{center}

Hydrogen selenide is the toxic, flammable selenium congener of more-familiar H$_2$O (water) and H$_2$S (hydrogen sulfide).  Though not corrosive to metals in cool and dry environments, like H$_2$S it degrades several common lab polymers\cite{mathelaChapter16Hydrogen2021}, including viton.  

We are interested in H$_2$Se in the gas phase only; it is polar, like water, and this {\em probably} prohibitively slows electron mobility in the liquid.  (Surprisingly, H$_2$S---and presumably H$_2$Se as well---does not share H$_2$O's large ionic conductivity\cite{wilkinsonLiquidHydrogenSulfide1931}.) 

First, do low-energy electrons attach to H$_2$Se to form H$_2$Se$^-$ anions?   DFT models of H$_2$Se and H$_2$Se$^-$ suggest the binding energy of H$_2$Se$^-$ is 0.93 eV higher than the binding energy of the neutral; this is a negative ``electron affinity'' which means the anion is unbound and will not form in the thermal-energy collisions associated with drift regions.  

Second, do low-energy electrons cause {\em dissociative attachment} to H$_2$Se, yielding an anion and a neutral?  Experimental data\cite{abouafLowEnergyElectron2008} shows anion appearance at electron beam energies above around 1~eV.  Agreeing with this, our DFT calculations show that all possible dissociation patterns require energy input, although formation of H$_2$+Se$^-$ requires the least: only 0.18 eV.  This is small enough to be somewhat concerning but not a showstopper.

Although there is no precedent for running either H$_2$Se or H$_2$S as a drift or gain gas, but there is precedent for H$_2$O\cite{deurquijoElectronDriftVelocities2014a}.  Water (for which we calculate an affinity of -1.23 eV) is routinely included in drift gases\cite{fribertImpactTraceAmounts2026,boyarski2003} at the 0.1\% level.  High-H$_2$O proportional counters have also been tested \cite{hukElectronAttachmentOxygen1988a, aneeshaavasthiPhDThesisPreparation2026} and seen drift and gain.

%\begin{figure}
%  \includegraphics[width=0.6\textwidth]{glovebox_photo.jpg}
%  \includegraphics[width=0.4\textwidth]{all_orig_overlay.pdf}  
%  \caption{(a) The CWRU toxic drift gas R\&D system (in commissioning). A glovebox will allow low-risk gas handling, with electrical and high-voltage feedthroughs to bring typical gas-detector electrical signals into the laboratory space.  It currently contains a proportional counter built with bakeable materials in preparation for running corrosive fluids above 100~C, as well as glassware for initial operations with hydrogen selenide and selenophene.  (b) Proportional counter checkout data showing $^{55}$Fe x-ray peaks in Ar:CH$_4$:H$_2$O 79:9:12 at 50C.}\label{fig_55fe}
%  \end{figure}

\subsubsubsection{CSeO}

\begin{center}
\chemname{\chemfig{Se=C=O}}{Carbonyl selenide\\CSeO\\m.p.\ -122, b.p.\ -23 C}
\end{center}

Carbonyl selenide (CSeO) is rarely studied or used; literature reports that it readily decomposes\cite{purcell217KineticsDecomposition1937} to Se and CO at elevated temperatures (100~C).  Here, in the absence of quantitative kinetics data, we are assuming it can be stored without decomposition at lower temperatures.  It is colorless polar liquid which boils to a colorless ``evil-smelling'' gas \cite{pearson83CarbonylSelenide1932a}.  Its laboratory preparation is fairly straightforward \cite{kondoNewSynthesisCarbonyl1979}. 

It is in the family of CO$_2$ congener molecules with oxygen substitutions (symmetric CO$_2$, CS$_2$, CSe$_2$ or asymmetric CSO, CSeO, CSeS).  This is interesting because CO$_2$ is familiar electron-drift medium and CS$_2$ is a familiar negative-ion-drift medium.  Where do the asymmetric molecules go?  After verifying that our DFT reproduces known CO$_2$ and CS$_2$ electron affinities, we predict an 0.28 eV threshhold for forming CSeO$^-$ from CSeO, implying that CSeO is electropositive and an electron drift medium.  Other DFT work agrees\cite{gutsevElectronAffinitiesCO21998a}.  In this simple system we were also able to run a full set of dissociative attachment reactions and find positive energy requirements for all of them, consistent with literature reports on CSO and CO$_2$\cite{hubin-franskinDissociativeElectronAttachment1976,macneilNegativeIonFormation1969}. 

These numbers are, however, well within the uncertainty of our DFT calculations.  If CSeO$^-$ turns out to be a very weakly bound, it may make CSeO interesting as a ``detachable'' negative ion drift species.  An in-between case---an electron affinity small enough to allow attachment/detachment by thermal fluctuations---would put CSeO in an unprecedented category worthy of further study.  

DFT predicts carbon diselenide (CSe$_2$) and thiocarbonyl selenide (CSeS) to be electronegative.

\subsubsubsection{Methaneselenol}

\begin{center}
  \chemname{\chemfig{Se([:-30]-H)([:-180]-C([:100]-H)([:190]<:H)([:210]<H))}}{Methaneselenol\\CH$_3$SeH\\m.p.\ -54 C, b.p.\ 12 C}
  
\end{center}Methaneselenol is the selenium congener of methanol and methanethiol (also known as methyl mercaptan, the warning odorant commonly added to natural gas).  We note that methanol has been used in drift chambers without attachment\cite{hukElectronAttachmentOxygen1988a} and decent gain performance\cite{brunnerTownsendCoefficientsGases1978}, and its dissociative electron attachment behavior is known\cite{thynneIonisationDissociationElectron1973} with a high threshhold observed.   Our DFT calculations predict that methaneselenol is electropositive with an EA of -0.67 eV.  

\subsubsubsection{Selenophene, selenolane, and benzeneselenol}

\begin{center}
  \chemnamewide{\chemfig{[:-30]Se*5(-=-=-)}}{\phantom{xx}Selenophene\phantom{xx}}{C$_4$H$_4$Se\\m.p -38 C, b.p.\ 110 C} 
  \chemnamewide{\chemfig{[:-30]Se*5(-----)}}{\phantom{XXX}Selenolane\phantom{XXX}}{C$_4$H$_8$Se\\m.p.\ ?, b.p.\ 135 C} 
  \chemnamewide{\chemfig{              SeH% 7
    -[:180,,1]% 4
      =_[:240]% 3
       -[:180]% 2
      =_[:120]% 1
        -[:60]% 6
            =_% 5
                 (
           -[:300]% -> 4
           )}}{\phantom{XXX}Benzeneselenol\phantom{XX}}{C$_6$H$_5$SeH\\m.p. ?, b.p. 183 C} 
\end{center}

Selenophene is a heterocyclic compound with four C and one Se in a five-member ring.  Its oxygen and sulfur congeners are furan (C$_4$H$_4$O) and thiophene (C$_4$H$_4$S); tellurophene (C$_4$H$_4$Te) and tellurolane (C$_4$H$_8$Te) will be discussed later.  It is a colorless to light yellow liquid with a vapor pressure of 46.5 torr at 25C.  Selenolane (AKA tetrahydroselenophene, C$_4$H$_8$Se) is a fully saturated heterocycle with the same skeleton\cite{morgan1929}; its congeners are tetrahydrofuran and tetrahydrothiophene.   Their use in a gas TPC would require an unusual elevated temperature, but this does not present obviously insurmountable problems.   Selenolane might have a safety benefit over selenophene: a simple ice water bath {\em probably} brings it below its predicted\cite{Selenacyclopentane} 36~C flash point.

The oxygen and sulfur congeners are widely used as precursors in organic synthesis, including thiophene's use in organic semiconductors; this has generated a wide literature on synthesis, purification, and safety which is absent for the other compounds we are discussing.   In addition, low-energy electron interactions with small heterocycles, of interest to biologists interested in radiation damage to DNA, has generated relevant literature, including both theory\cite{swadiaElectrondrivenProcessesFuran2017} and experiment.  The thiophene anion (C$_4$H$_4$S)$^-$ forms at 1.046~eV above the neutral \cite{upadhyayaTheoreticalStudyDissociation2024}.  For its congener selenophene, DFT predicts EA=-0.63 eV.   Tetrahydrofuran (C$_4$H$_8$O) is observed to attach electrons of energy 1.25~eV and has also shown electron mobility \cite{kadhumReactivitySolvatedElectrons1986}.  For their congener selenolane, DFT predicts EA=-0.67 eV.

We did not exhaustively search for dissociative attachment pathways but feel they are unlikely due to their absence in thiophene and tetrahydrofuran.

Benzeneselenol (the selenium congener of phenol and thiophenol)\cite{bradtOrganicCompoundsSelenium1935a} is a yellow liquid whose foul odor is particularly notorious\cite{lowe_selenophenol}.  In thiophenol, electron beam data suggest an attachment energy of -0.7 to -1.1 eV, and drift experiments show a 0.7 eV onset for attachment (leading to dissociation to C$_6$H$_5$S$^-$); an attachment coefficient vs reduced electric field is given in \cite{spyrouElectronswarmMassspectrometricStudy1999}.  Using this to crosscheck our DFT, we give thiophenol a VEA of -0.4 eV to the intact ion and EA -1.39 eV to the fragments.  The same DFT for benzeneselenol gives a VEA=-0.51 eV (and -1.1 eV to fragment) suggesting it resembles thiophenol and is electropositive.

\subsubsubsection{Other selenium compounds}

\begin{center}
  \chemnamewide{\chemfig{          
      [:-30]Se*6(---Se---)}
    }{1,4-diselenane}{C$_4$H$_8$Se$_2$\\m.p.\ 113 C, b.p. 237? C}
    \phantom{XXXXXX}
  \chemnamewide{\chemfig{            HSe% 7
      -[,,2]% 6
    =^[:300]% 5
           -% 4
     =^[:60]% 3
               (
         -[,,,1]SeH% 8
               )
     -[:120]% 2
    =^[:180]% 1
               (
         -[:240]% -> 6
         )}}{Benzene 1,4-diselenol}{C$_6$H$_6$Se$_2$\\m.p~$>$97C, b.p.\ ? C}
\end{center}

None of the electropositive gases found so far have been {\em nonpolar}, a particularly desirable property for a two-phase TPC.  Although liquid phase compatibility is not the focus of this paper, we briefly asked whether any nonpolar compounds could be identified for possible use in the liquid phase.  The candidates are two-selenium molecules, possibly useful in the gas phase as well albeit at fairly high boiling points.  1,4-diselenane (reported\cite{gouldReactions14Diselenane11956} to be a white volatile solid melting at 113C; one online catalog reports a predicted boiling point of unknown origin and accuracy\cite{14Diselenane1538416Wiki}) is shown with DFT to have an EA of -0.58~eV.  (An unsaturated version, 1,4-diselenin (C$_4$H$_4$Se$_2$) seems plausible but unreported.  If it adopts the ``boat'' configuration of the sulfur congener 1,4-dithiin\cite{kobayashiChemistry14Dithiins1986} would, in any case, not be nonpolar.)

Benzene 1,4-diselenol (AKA 1,4-benzenediselenol), though not attested in the literature, seems plausible given the existence of benzene 1,4-dithiol (melting point 92--97 C; in later tables in this paper, we assign b.p. 250~C when a boiling point is called for.)).  DFT gives it a VEA of -0.36 eV.  

Selenium compounds for which we found positive EAs can be seen in Table~\ref{tab_en}.  Note that dimethyl selenide (CH$_3$SeCH$_3$) is a special case with a negative electron affinity with respect to the intact ion but where DFT suggests an energetically-favorable DEA pathway (CH$_3$SeCH$_3$ + e$^-$ $\rightarrow$ Se$^-$+ C$_2$H$_6$), similar to what is seen experimentally in dimethyl sulfide\cite{kopyraElectronDrivenProcesses2015}.

\subsubsection{Tellurium}

\begin{center}
\chemnamewide{\chemfig{[:-30]Te*5(-=-=-)}}{\phantom{xxxxxxx}Tellurophene\phantom{xxxxxxx}}{C$_4$H$_4$Te\\m.p.\ -36 C, b.p.\ 148 C}
\chemnamewide{\chemfig{[:-30]Te*5(-----)}}{\phantom{xxxxxxx}Tellurolane\phantom{xxxxxxx}}{C$_4$H$_8$Te\\m.p.\ ??, b.p.\ 166--167 C} 
\end{center}

$^{130}$Te ($Q_{\beta\beta}=2.528$ MeV, $f$=34.50\%) is the focus of this section although some readers may be interested in $^{128}$Te.  Tellurium, like selenium, is a copper-mining byproduct, with a market for 600T/y and valued around USD60/kg.  It is on the US Critical Materials List due to its limited supply and its use in CdTe solar cells.  The search for a tellurium-bearing TPC fluid could follow many of the same lines we followed for $^{82}$Se, with the caveat that the chalcogen binding grows less and less strong as we move down the periodic table.  For example, H$_2$Te and CH$_3$TeH are thermally unstable and disproportionate to metallic tellurium, even at liquid temperatures.  (Even if  H$_2$Te were usefully stable at lower temperatures, it is allowed to attach electrons by dissociation to H$_2$ + Te$^-$.)  CTeO and CTeS are reported to be unstable.

However, tellurophene\cite{mackSynthesisTelluropheneIts1966} and tellurolane\cite{morganXXIVcycloTellurobutaneTetrahydrotellurophen1931} (AKA tetrahydrotellurophene) are reportedly stable and our DFT modeling shows them to be electropositive.  We have not attempted to model all dissociative attachment pathways.  

A DFT survey of halogenated telluroformaldehydes (TeCH$_2$) predicts some may be stable but all would be electronegative\cite{jaufeerallyTelluroformaldehydeItsDerivatives2012}.

  Teflic acid (HOTeF$_5$) engages in interesting HF-like chemistry\cite{baderPentafluoroorthotelluratesRelatedCompounds2025a} to bond to transition metals, with many of the results being volatile and surprisingly stable; B(OTeF$_5$)$_3$ (boron pentafluoroorthotellurate or boron teflate) is (like BF$_3$) a volatile nonpolar substance.  Our survey has not revealed an electropositive volatile teflate yet but this merits further study.  
  
  \subsubsection{Germanium}

\begin{center}
\chemnamewide{\chemfig{Ge([:90]-H)([:200]-H)([:10]<:H)([:-30]<H)}}{\phantom{xxxx}Germane\phantom{xxxx}}{GeH$_4$\\ m.p -165 C, b.p.\ -88 C}
\chemnamewide{\chemfig{Ge([:90]-C)([:200]-C)([:10]<:C)([:-30]<C)}}{Tetramethylgermanium}{Ge(CH$_3$)$_4$\\m.p.\ -88 C, b.p.\ 43 C}
\end{center}

$^{76}$Ge ($Q_{\beta\beta}=2.039$ MeV, $f$=7.80\%) is one of the best-studied \ovbb~targets thanks to its extraordinary performance in semiconductor diodes (largely overcoming the disadvantages of a low $Q_{\beta\beta}$).  However, semiconductor-based $^{76}$Ge \ovbb~searches may not scale well beyond the ton scale.   Germanium is produced in quantities of 300-400 T/y and valued at USD1000/kg; this is expensive but supply is probably more elastic than it is for xenon.  It is on the US Critical Materials List due to its use in semiconductors and optical glasses.
  
Two germanium compounds, germane and tetramethylgermanium, have precedent as ionization counter fluids\cite{holroyd1991,masudaCharacteristicsIonizationChamber1993} with electron drift in the liquid phase; therefore are almost certainly viable drift media in the gas phase; the congener tet  DFT affirms negative electron attachment energy for both.

In contrast to CF$_4$, a widely-used electropositive (and scintillating) drift gas, germanium tetrafluoride is electronegative. (Since CCl$_4$ is electronegative\cite{suElectronAttachmentRate2012}, the electronegativity of GeCl$_4$ is unsurprising.)  We believe that the mono- and di-halogenated germanes GeH$_3$[X] and GeH$_2$[X]$_2$ and various methylgermanes Ge(CH$_3$)$_n$[X]$_{4-n}$\cite{buschInorganicSynthesesVolume1978} are unstable but this is not certain.  We have not comprehensively surveyed the heavier or mixed halides, or other organogermanium compounds.  However, it seems plausible that compounds with specific desired properties---lower toxicity, possibly scintillation---could be sought, albeit not in the smallest/lowest-boiling species. 

  \subsubsection{Tin}
  
\begin{center}
\chemnamewide{\chemfig{Sn([:90]-H)([:200]-H)([:10]<:H)([:-30]<H)}}{\phantom{xxxxxxxx}Stannane\phantom{xxcxxxxxx}}{SnH$_4$\\ m.p.\ -146 C, b.p.\ -52 C}
\chemnamewide{\chemfig{Sn([:90]-C)([:200]-C)([:10]<:C)([:-30]<C)}}{Tetramethyl tin}{Sn(CH$_3$)$_4$\\m.p -54 C, b.p.\ 75}
\end{center}

  $^{124}$Sn (endpoint 2.287 MeV, $f$=5.80\%) is an under-studied \ovbb~candidate; the \tvbb~process has not yet been observed.  Tin is common, inexpensive (USD20/kg), and surprisingly radiopure\cite{agrawalRadioassayMaterialsAMoREII2024a}. The most volatile tin compound is the methane/silane/germane congener stannane (SnH$_4$)\cite{garzaStannaneExtremeUltraviolet2023}, which DFT calculations show to be electropositive with electron affinity of -0.9 eV.  At first glance, this is no use: stannane is tabulated as ``thermally unstable'' and decomposing to Sn(s) + 2H$_2$ at room temperature, which seems like a showstopper.  However, from the literature \cite{tamaruThermalDecompositionTin1956} we get interesting details.  The decomposition is a surface effect, occurring on metallic tin and not on tin oxide or glass.  From the reported first-order kinetics, and the deposition rate shown at 35C, a tin surface under SnH$_4$ appears to accumulate tin at 50~$\mu$g/cm$^2$/h/atm, which extrapolates to 140~ng/cm$^2$/h/atm at -50 C.  This is slow enough that it is not really a stannane {\em storage} problem, although it might be a stannane detector functioning problem.  It is not clear how long an initially-untinned surface would survive.   Interestingly, stannane in a SnH$_4$-O$_2$ mixture was reported not to decompose, since the oxygen quickly passivated any newly-deposited tin and prevented the surface catalysis.  Could a self-passivating drift gas mixture (not oxygen), or a regenerable frost layer, or some other intervention, protect initially-untinned surfaces in a large stannane TPC?  

The next-most-volatile tin compound is thermally stable: Tetramethyltin (Sn(CH$_3$)$_4$), which appears to be electropositive (vertical EA=0.67 eV) in DFT, which agrees with a measured electron mobility and ion yields in the liquid state\cite{schmidtElectronMobilityNonpolar1977,holroyd1991}.  Mobility appears to be comparable to that of tetramethylsilane and tetramethylgermane.   Tetraethyltin (b.p.\ 181 C) and tetrabutyltin (b.p.\ 245 C) electron mobilities have also been surveyed\cite{holroyd1991}, but these seem less practical in the gas phase.

Tin tetrafluoride has a very high melting point and tin tetrachloride is electronegative.
  
\subsubsection{Zirconium}

\begin{center}
  \setchemfig{atom sep=2em}
    \chemnamewide{\chemfig{Zr(-[:180]N(-[:120])-[:240])(-[:270]N(-[:330])-[:210])(-[:90]N(-[:150])%
-[:30])-N(-[:300])-[:60]}}{Tetrakis(dimethylamino)Zr(IV)}{((CH$_3$)$_2$N)$_4$Zr\\m.p.\ 60 C, b.p.\ $\sim$89 C}
\chemnamewide{\chemfig{Zr(-[:225]O-[:225](-[:135])(-[:225])(-[:315]))
  (-[:135]O-[:135](-[:135])(-[:45])-[:225])
  (-[:315]O-[:315](-[:45])(-[:315])-[:225])
  -[:45]O-[:45](-[:45])(-[:135])-[:315]}}{Zr(IV) tert-butoxide}
  { \phantom{XX} Zr(OC(CH$_3$)$_3$)$_4$ \phantom{XX}\\
     m.p.\ 3 C , b.p.\ 250 C }
\chemnamewide{
\chemfig{
  Zr(-[:210]O-[:270]-[:210])
             (
        -[:120]O% 5
       -[:180]% 6
        -[:120]% 7
             )
             (
       -[:300]O% 8
       -[:360]% 9
       -[:300]% 10
             )
    -[:30]O% 11
    -[:90]% 12
    -[:30]% 13
}}{\phantom{XX}Zr(IV)ethoxide\phantom{XX}}{Zr(OC$_2$H$_5$)$_4$\\m.p.\ $\sim$172 C, b.p.\ $>$235 C}
\end{center}

$^{96}$Zr ($Q_{\beta\beta}=3.349$ MeV, $f$=2.80\%) is an under-studied \ovbb~nuclide with an unusually high endpoint (3.349 MeV).  Zr is a common metal, mined in quantities near 1MT/y and valued around USD40/kg.  It is on the US Critical Materials List due to a lack of domestic refining capacity rather than to scarcity.  Its not-yet-observed single beta decay\cite{finchSearchDecay96Zr2016} (Q = 119 keV to $^{96}$Nb$(5^+)$ followed by $\beta/\gamma$ cascade to $^{96}$Mo) would be an interesting secondary physics target.

Zirconium's only known volatile inorganic compound is Tetrakis(tetrahydroborate)zirconium (Zr(BH$_4$)$_4$) with a reported 32~C melting point an an extrapolated boiling point of 132~C.  However, it is thermally unstable even around 32~C\cite{hoekstraPreparationPropertiesGroup1949a}, decomposing to diborane B$_2$H$_6$\cite{gennariSynthesisThermalStability2009} when the vapor pressure is only a few tens of torr.  At first glance this seems to preclude its use as a detector gas at large scale.  The decomposition reaction is reversible (hence interest in metal tetrahydroborates as hydrogen storage media); perhaps one can imagine a large TPC in which a Zr(BH$_4$)$_4$, B$_2$H$_6$, and H$_2$ gas mixture establishes an equilibrium including high density of gas-phase Zr.  H$_2$ (EA = -2.24 eV) and B$_2$H$_6$ (EA = -1.0 eV) are electropositive while our DFT model of Zr(BH$_4$)$_4$ is so weakly electronegative (VEA = 0.15 eV) that we cannot be confident in the sign.  (We are mentioning this mostly for completeness; it sounds implausible and we will exclude it from further tables.)

To find Zr compounds with accessible liquid and gas phases, we turn to the metal-organic chemical vapor deposition (MOCVD) industry, which is interested in vapor-phase transport of metals to surfaces.  The focus of MOCVD research is on behavior at low pressures, so further work will be need to understand these compounds' physical properties in the conditions of a TPC.   We will attempt DFT models of the intact anions but have not modeled any possible dissociation pathways.  

We found three volatile Zr compounds which our DFT work labels as electropositive.  Interestingly, all are also nonpolar and might be worth investigating in the liquid state.

\paragraph{Tetrakis(dimethylamino)zirconium (Zr(NEt$_2$)$_4$ or TDMAZ)} TDMAZ melts at 60~C and forms a gas-phase monomer with 1~torr vapor pressure at 80~C.  Published vapor pressure data \cite{vikulovaStudyThermalProperties2024} suggest we reach 751~torr at a ``recommended holding temperature'' of 89~C, but we do not know the decomposition kinetics.  DFT suggests TDMAZ has a electron affinity of -0.75 eV.  

\paragraph{Zr(IV) tert-butoxide (Zr(O{\em t}Bu)$_4$ or ZTBO)}  Zr(O{\em t}Bu)$_4$ is a light yellow liquid at room temperature and is stable enough to distill at 1~bar and 250C\cite{bradleyPyrolysisMetalAlkoxides1959}, although accurate vapor pressure data only exists up to 166C\cite{bradley150VapourPressures1959}.  It may polymerize under cold storage \cite{spijksmaZirconiumHafniumTertbutoxides2013}.  DFT suggests ZTBO has a vertical electron affinity of -0.7 eV. 

\paragraph{Zirconium(IV) ethoxide (Zr(OEt)$_4$)} Zr(OEt)$_4$\cite{bradley62ZirconiumAlkoxides1951} appears electropositive with a VEA of -0.75 eV, but its high temperature scale ($\sim$170~C melting, 5~torr vapor at 235~C; in later tables in this paper, we assign b.p. 250~C when a boiling point is needed.) seems less favorable for our needs.

\subsubsection{Molybdenum}

\begin{center}
  \chemnamewide{
    \chemfig[angle increment=30]{((-[,1.41,,,draw=none](-[9,,,,style=dashed]Mo-[9,,,,style=dashed]-[6,1.41,,,draw=none](-[6]-[7])=[1]-=[11]-[7]=[6]-[5]))=[1]-=[11](-[0]-[1])-[7]=[6]-[5])}
  }
               {Bis(ethylbenzene)molybdenum}{(C$_{8}$H$_{10}$)$_2$Mo\\m.p.\ $<$23 C. b.p.\ 150--170 C (1 torr)}
\end{center}
  
 $^{100}$Mo ($Q_{\beta\beta}=3.034$ MeV, $f$=9.60\%) is a heavy isotope of an abundant and inexpensive transition metal valued around USD40/kg. It is one of the handful of transition metals with volatile fluorides and carbonyls.  Unfortunately, MoF$_6$ (molybdenum hexafluoride) and Mo(CO)$_6$ (molybdenum hexacarbonyl)\cite{monteVolatilityChemicalStability2018} are known to be electronegative\cite{miyoshiElectronAffinitiesHexafluorides1988,shihLowenergyElectronInteraction2022}. Our DFT modeling also finds electronegative EAs for several obscure oxyfluorides.  MoF$_5$ is also volatile, with various oligomers appearing in the gas phase; DFT models of the monomer and the (probably dominant) tetramer show them as electronegative.  We believe our survey has exhausted the volatile small molecules, so we turn to complexes.

We used DFT to analyze two low-melting-point complexes available as MOCVD precursors, including Bis(ethylbenzene)molybdenum (Mo(EtBz)$_2$).  Mo(EtBz)$_2$, an example of a diarene ``sandwich'' compound, is reported to be a ``dark green liquid''\cite{BISETHYLBENZENEMOLYBDENUM32877002}.  DFT finds it to be electropositive with a vertical electron affinity of -0.76~eV.  Dissociative attachment of the form (EtBz)$_2$Mo + e$^-$ $\rightarrow$ 2 EtBz + Mo$^-$  is also disfavored; we did not check other pathways.  Pyrolysis data and decomposition kinetics are available from higher temperature measurements\cite{dyagilevaKineticStabilityBiscyclopentadienyl1988} (340--400~C) suggesting reasonable stability at lower temperatures, and a propensity to decompose on surfaces rather than in the gas phase.  (In later tables in this paper, we assign b.p. 250~C when a boiling point is called for.)  Since it is a nonpolar molecule, readers may be interested in its liquid phase as well as its gas phase.

It would be interesting to study other other molybdenum diarenes.

\subsubsection{Calcium, palladium, neodymium, and cadmium}

We surveyed some \ovbb~nuclides for which we failed to find electron drift gases for various reasons.  We did not consider $^{238}$U, various $Q_{\beta\beta}<2$~MeV nuclides\cite{suhonenDoublebetaDecays70Zn2011}, or double electron capture candidates.

  \paragraph{Calcium} $^{48}$Ca ($Q_{\beta\beta}=4.267$ MeV, $f$=0.19\%) is notable for its high $Q_{\beta\beta}$.  This energy, far beyond most beta/gamma backgrounds, seems like it might allow detector construction to focus on extreme size at the expense of background reduction.  As we will see later, the low isotope abundance and unfavorable matrix elements make this hard to pull off in a TPC even if a fluid were available.  In any case, there seems to be no easy way in to the chemistry.  Somewhat volatile MOCVD precursors include calcium acetylacetonate (melting point reported $\sim$165 C), calcium bis(cyclopentadienyl) or calcocene (melting point unknown, but the magnesium congener melts at 176C), and Ca(2,2-dimethyl-3,6,9-trioxa-2-siladecane)(hexafluoroacetylacetonate)$_2$ (melting at 70C \cite{parkSilylationDrivenVolatilityEnhancement2024} but very large).   We have not attempted DFT attachment-energy calculations.

  \paragraph{Palladium}
  $^{110}$Pd ($Q_{\beta\beta}=2.017$ MeV, $f$=11.80\%) is one of the less-studied \ovbb~candidates, in part due to its low endpoint (2.017 MeV).  For beyond-ton-scale science, its precious-metal-level cost (above USD~50,000/kg) and global supply limitations appear worse than xenon's.  At modest scales, maybe for \tvbb~science, it may be worth revisiting the cost comparison in upcoming decades as internal combustion engines, whose catalytic converters drive most palladium demand, become obsolete.  Despite nickel's stable carbonyl Ni(CO)$_4$, a corresponding palladium(0) carbonyl Pd(CO)$_4$ does not exist\cite{stromnova1998} and (unless the unidentified complex carbonyl reported by Libuda\cite{libuda1995} could be identified) we have found no other inorganic compounds of interest. Interesting Pd compounds will therefore be complexes, the most readily available of which is cyclopentadienyl allyl palladium (melting point around 60C, vapor behavior unknown).  We have not found literature reports of EAs and we have not attempted DFT attachment-energy calculations.
  
  \paragraph{Neodymium} 
 $^{150}$Nd ($Q_{\beta\beta}=3.371$ MeV, $f$=5.60\%) is attractive as a high-Q \ovbb~search target in a mid-priced (USD~60/kg) element, but we find no compounds worth consideration as TPC gases.  Available MOCVD precursors for Nd deposition\cite{strem} are very large complexes, like tris(hexafluoroacetylacetonato)neodymium diglyme (899 amu, m.p.\ 73--75 C)\cite{nigro2009} or neodymium tris(hexafluoroacetylacetonate) (765.39 amu, m.p.\ 143 C, b.p.\ $>200$ C)\cite{strem} with discouragingly high melting points.   (A size/volatility anticorrelation seems to be systematically true of lanthanide complexes\cite{drozdovVolatileCompoundsLanthanides2013}; the central metal has a large ionic radius and can only be screened by very bulky ligands.)  Literature reports show zero-energy nondissociative and dissociative electron attachment to metal hexafluoroacetylacetonate complexes with Cu and Pd\cite{engmannDissociativeElectronAttachment2013}, as well as to the ligands themselves\cite{landheerLowEnergyElectronInducedDecomposition2011}.   We have not attempted DFT calculations.
  
  \paragraph{Cadmium} 
$^{116}$Cd ($Q_{\beta\beta}=2.813$ MeV, $f$=7.60\%) we treat as a special case.  Cadmium is one of the least-expensive elements (under USD~3/kg) because it's a byproduct of zinc mining that nobody wants.  Cadmium, like zinc and mercury, has rich organometallic chemistry including volatile species like diethyl cadmium and dimethyl cadmium.  Even by the standards of this paper, which is already a catalog of toxic and hazardous compounds, organocadmium compounds present more daunting safety hazards than this author wishes to deal with\cite{ThingsWontWorka} even at gram scales, much less to invite a research community to deal with at ton scale.  For this reason alone, we have ignored them in our survey.

\subsection{Candidate electronegative gases}

In searching for electropositive gases, we surveyed a selection (but by no means a comprehensive one) of other compounds with near-room-temperature gas and liquid phases.  In the process we came across many electronegative compounds, listed in Table \ref{tab_en}.  These might find uses in negative-ion TPCs.  TPCs whose mobile charge is an anion, rather than an electron, are valuable because the ions diffuse very little during drift, allowing high-resolution track imaging\cite{phanNovelPropertiesSF62017,ohareRecoilImagingDirectional2022,battatDarkMatterTime2014}.  This approach, familiar in nuclear-recoil-detection contexts, is under study\cite{nygrenNeutrinolessDoubleBeta2018} and/or small-scale implementation in SeF$_6$\cite{nvdex} using the TOPMETAL ion sensing architecture.  If this works, an ion-sensing TPC could serve as a multi-isotope \ovbb~observatory platform.  

Negative ion TPCs are not {\em absolutely} dependent on direct ion sensing.  CS$_2$, which has a low electron affinity (0.16~eV in our DFT, higher in literature) is known to detach electrons (probably collisionally) in high electric fields and at low pressures, allowing electron avalanches to begin\cite{dionMechanismTownsendAvalanche2010} and enabling conventional gain readout of ion-drift detectors\cite{battatDarkMatterTime2014}.  Some of our gases (CSe$_2$, CSSe, SeCF$_2$, SeCH$_2$, Te(OH)$_6$, Cycloheptatriene molybdenum tricarbonyl) we identify as {\em fairly weakly} electronegative; given our uncertainties, they may turn out to be slightly electropositive, or have CS$_2$-like detachability.  Can we strip electrons from these anions?  At what pressures?  Can we use laser detachment\cite{sowadaLaserPhotodetachmentElectrons2008}?  

\begin{table}
      \begin{center}{\bf Negative ion drift \ovbb~TPC gas candidates}\end{center}
\begin{tabular}{r@{\hspace{-.5cm}} r p{3.5cm}  r r r p{3cm} }
&   &   & m.p. & b.p. & EA \\
&      Formula    & Name       & (C) & (C) & (eV) \\  
  \hline
&  CS$_2$ & Carbon disulfide & & & 0.167 & vs 0.5--0.8\cite{gutsevElectronAffinitiesCO21998a}\\
&  O$_2$ & Oxygen & & & 0.86 & vs 0.45\cite{schiedt1995}\\
    \hline
    \multirow{2}{*}{$^{86}$Ge} & GeF$_4$ & Ge tetrafluoride & \multicolumn{2}{c}{(t.p.\ -15 C)} & 1.47 & \\
     & GeCl$_4$ & Ge tetrachloride & -49 & 86 & 1.0 & \\

  \hline
\multirow{11}{*}{$^{82}$Se} &  CSe$_2$ & Carbon diselenide & -44 & 125 & 0.57 \\
&  CSSe & Thiocarbonyl selenide & -75 & 84 & 0.38 \\
&  SeF$_4$ & Selenium tetrafluoride & -13 & 101 & 1.8 & \\
&  SeF$_6$ & Selenium hexafluoride & -39 & -34 & 3.32 & also DEA\cite{nygrenNeutrinolessDoubleBeta2018} \\
&  SeCF$_2$ & Selenocarbonyl fluoride & -81 & 45 &  0.68 \\
&  SeCH$_2$ & Selenoformaldehyde & ? & ?  & 0.82 \\
&  SeOF$_2$ & Seleninyl fluoride & ?  & 125 & 1.4 \\
&  SeO$_2$F$_2$ & Selenoyl fluoride & -99 & 8 & 3.3   \\
&  SeF$_5$OF & Pentafluoroselenium hypofluorite & -54 & -29 & 3.1\rlap{$^V$} \\ 
&  SeOCl$_2$ & Seleninyl chloride & 11 & 180 & 2.93  \\
&  CH$_3$SeCH$_3$ & Dimethyl selenide & -87 & 55 & 0.66 & also DEA\cite{kopyraElectronDrivenProcesses2015} \\
&  CH$_3$Se$_2$CH$_3$ & Dimethyl diselenide & ?? & 157 & -0.77 & DEA$^*$\cite{kopyraElectronDrivenProcesses2015} \\
&  C$_{12}$H$_{10}$Se$_2$ & Diphenyl diselenide & 60 & $>$202 & ? & DEA$^*$\cite{modelliTemporaryAnionStates2006a}\\
\hline
\multirow{1}{*}{$^{96}$Zr}
&  Zr(BH$_4$)$_4$ & Zr tetraborohydrate & 32 & ? & 0.15\rlap{$^V$}\\
\hline
\multirow{7}{*}{$^{100}$Mo} &  Mo(CO)$_6$ & Mo hexacarbonyl & 150 & 156 & 0.5 & also DEA\cite{shihLowenergyElectronInteraction2022} \\
&  MoF$_6$ & Mo hexafluoride & 17 & 34 & 3.9 & also DEA\cite{friedmanElectronAttachmentMoF62006b}  \\
& MoO$_2$F$_2$ & Mo difluoride dioxide & \multicolumn{2}{c}{(subl 270 C)} & 2.1 & \\
& MoO$_2$Cl$_2$ & Mo dichloride dioxide & 175 & ? & 2.4 & \\ 
&  MoOF$_4$ & Mo oxytetrafluoride & 95 & 186  & 3.28 \\
&  MoF$_5$ & Mo pentafluoride & 45 & ?? & $>$5 & (tetramer)\cite{steneMoF5RevisitedComprehensive2018a} \\
& C$_7$H$_8$Mo(CO)$_3$ & Cycloheptatriene~molybdenum tricarbonyl & $\sim$100 & ? & 0.7\rlap{$^V$}\\
\hline
\multirow{3}{*}{$^{124}$Sn}
&  Sn(NO$_3$)$_4$ & Tin(IV) nitrate & 91 & 98  & 3.63 & \\
&  SnCl$_4$ & Tin(IV) chloride & -34 & 114 & 2.8  & \\
&  SnBr$_4$ & Tin(IV) bromide & 31 & 205 & 2.23\rlap{$^V$}   & \\
\hline
\multirow{4}{*}{$^{130}$Te}
& TeF$_6$ & Te hexafluoride & -39 & -38 & 2.3\rlap{$^V$} & also DEA\cite{jarvisInvestigationsLowEnergy2001} \\
& HOTeF$_5$ & Teflic acid & 39 & 60 & 2.9\rlap{$^V$} \\
& Te(OH)$_6$ & Orthotelluric acid & 136 & ? & 0.54\rlap{$^V$} \\
& B(OTeF$_5$)$_3$ & Boron teflate & ? & ? & 3.2\rlap{$^V$}\\
\hline
\end{tabular}
\caption{Volatile but electronegative \ovbb~compounds identified in our DFT and literature survey.  EAs with superscript$^V$ are vertical electron affinities, others are adiabatic.   Asterisks$^*$ indicate where DEA evidence is from a congener rather than the compound of interest.  We include two familiar electron-attaching gases, CS$_2$ and O$_2$ as a sort of calibration, showing DFT/experiment agreements at the level of $\approx$0.5 eV.  Note that DFT errors of this magnitude could flip the signs of some electron affinity assignments. }\label{tab_en}
\end{table}

\section{Large-scale gas TPCs and gas selection criteria}

Any future gas TPC for a \ovbb~search needs to have (a) a large mass of the target isotope, such that the event rate is high, and (b) event topology reconstruction good enough that the background rate is low.  In gas TPCs these two criteria are closely related and need to be considered together.  However, economic considerations usually ask us to minimize {\em something}, sometimes channel count or instrumented volume, in a way consistent with maintaining sensitivity.  In this section, we attempt to set up the problem of large-detector design optimization in a way that lets us compare the gas options listed in Table~\ref{tab_electropositive}.  

\subsection{Stopping power, tangling power, and defining a gas figure-of-merit}
\subsubsection{Detector design around stopping power}

Gas-phase TPCs will rely on {\em track imaging} to distinguish single from double beta particles\cite{iguaz2017,irastorza2016}, particularly for single beta particles at the \ovbb~endpoint energy.  Discrimination is only possible if the detector spatial resolution scale $\sigma_x$ is much smaller than the track length scale $L(E)$ at the energy of interest.  $\sigma_x$ might be determined by the density of anode plane instrumentation, or by the length of drift cells in which diffusion proceeds.  All else being equal, we will want to settle for the largest tolerable $\sigma_x$.

The track length $L(E)$ is determined by the stopping power of the material and by the density.  The density depends on both the gas choice and the operating temperature and pressure.  All else being equal, we would like to build smaller detectors rather than larger ones, driving us towards the highest tolerable density and the shortest tolerable $L(E)$.

Combining these two considerations, $L(E)/\sigma_x$ is a dimensionless number that corresponds to ``the number of detector resolution elements along a track''.  A given detector design problem might set a {\em floor} to $L(E)/\sigma_x$; you might make a statement like, for example, ``$L(E)/\sigma_x < 100$ does not allow adequate background rejection''.  Thereafter, designing an {\em affordable} detector (minimizing volume and/or minimizing instrumentation) requires working close to this floor.

\subsubsection{Track tangling through large-angle scatters}
A stopping power figure-of-merit might suffice when you have a known gas and only have to choose its operating pressure and readout details.  There is an additional complexity when we are comparing different gases: tracks look different in different gases due to different scattering angular distributions off low-Z and high-Z atoms.   A straight track is easier to reconstruct than a spaghetti-like tangle of the same length.  To provide simple, general guidance as to the extent of tangling, we wish to define the {\em tangling power} of a gas to measure its propensity towards large-angle scatters that add up into kinked or spaghettified tracks.

The radiation length $X_0$ is a good proxy for the high-Z, angular-deviation power of a scattering medium.  The most familiar formula\cite{tsai1974} for RMS angular deviation for a $z=1$ particle crossing a small distance $x$ is: 

\[
\theta = \frac{13.6 \mathrm{MeV}}{\beta c p}\sqrt{\frac{x}{X_0}}(1 + \mathcal{O}\left(\log\left(\frac{x}{X_0}\right)\right)
\]

We define a simple dimensionless quantity $\theta_t$, which we call the {\em tangling measure}, for a track coming from initial energy $E$ to a stop over a large distance $L$:  

\[
\theta_t \equiv \frac{27.2 \mathrm{MeV}}{E}\sqrt{\frac{L}{X_0}}
\]

\begin{figure}
  \begin{center}
    \includegraphics[width=\textwidth]{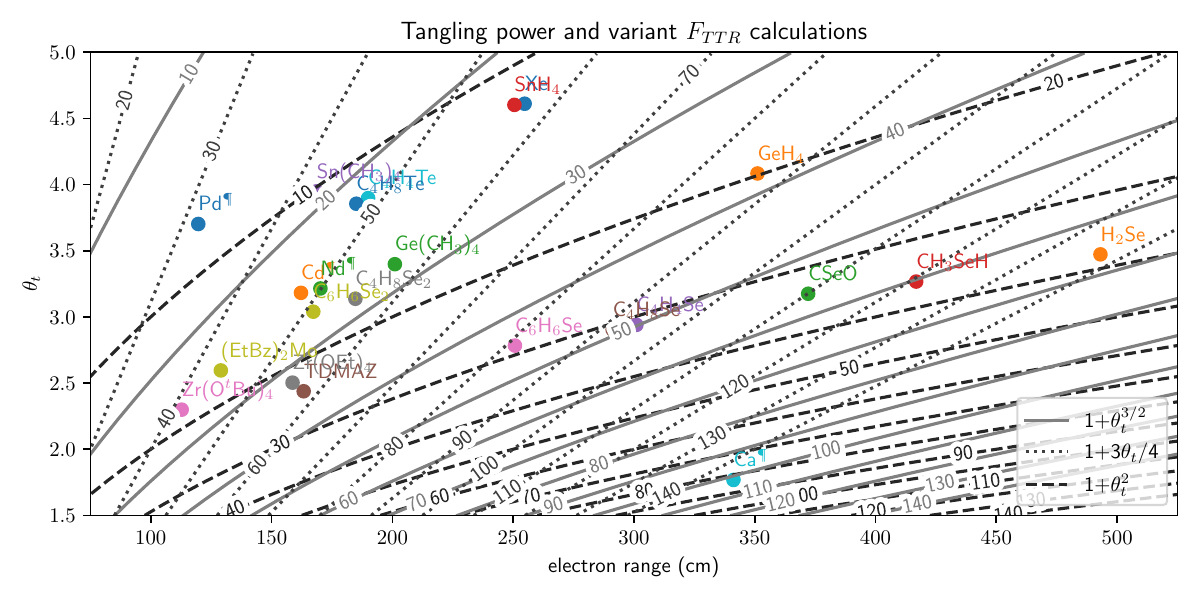}
    \end{center}
  \caption{The ingredients of the tangling-factor calculation and the evaluation of $F_{TTR}$ under different $f(\theta_t)$ models: the strawman $f(\theta_t) = 1+\theta_t^{3/2}$ formula used elsewhere in the paper, plus two reasonable-looking variants (further shown in Figure \ref{fig_FOM_variant}).  The track lengths are given in cm under a constant 44.6 mol/m$^3$ molar density.  Grey lines show contours of constant $F_{TTR}$ in the steps of 10~cm, for one guess involving stronger dependence  $f(\theta_t)=1+\theta_t^2$ and one with weaker dependence  $f(\theta_t)=1+3\theta_t/4$. (In a ``$\theta_t$ does not matter'' scenario, not plotted, the contours would be vertical.)  With some effort the reader can read off the impact of different assumptions.  For an example of a gas-selection impact, in the weaker-dependence scenario, for example, H$_2$Se has a nearly-factor-of-two advantage over C$_6$H$_6$Se, while in the strongest-dependence scenario the advantage is closer to 30\%.  For an example of an isotope-selection impact, (EtBz)$_2$Mo underperforms xenon in the weak-dependence scenario but outperforms it in the strong-dependence scenario.}\label{fig_tangling_var}
  \end{figure}

Some concrete examples of tracks with different tangling measures are shown in Figure~\ref{fig_tangling_var}.

\subsubsection{Defining figures-of-merit for detector sensitivity and discovery potential}
\begin{figure}
  \begin{center}
    \includegraphics[width=0.6\textwidth]{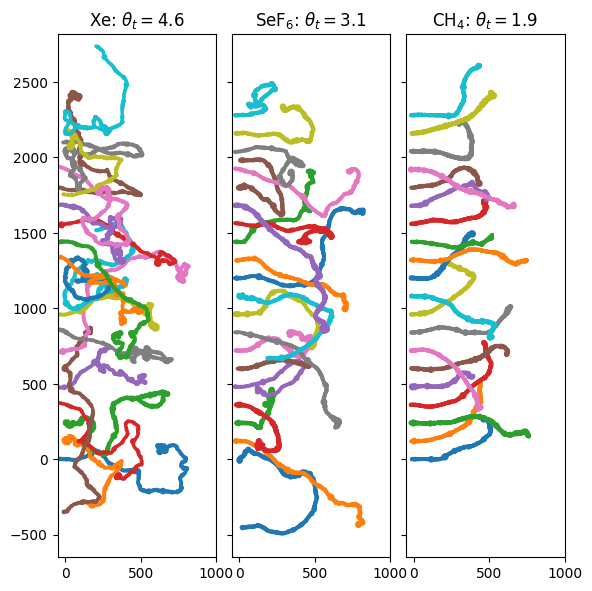}
    \end{center}
  \caption{Illustration of the meaning of ``tangling power'' $\theta_t$.  The figure shows 2D projections of the tracks of stopping 2.5 MeV electrons, initially moving in the x-direction from starting points placed evenly in y.  The electrons come to a stop in (a) Xe ($\theta_t=4.6$) (b) SeF$_6$ ($\theta_t=3.1$), or (c) CH$_4$ ($\theta_t=1.9$.), with pressures chosen such that all have the same mean track length.  Simulations are done with the Geant4-based package NEXUS\cite{nexus}.  It is qualitatively clear that, all else being equal, the lower-$\theta_t$ tracks will allow easier and less-error-prone track reconstruction.}\label{fig_tangling}
  \end{figure}

To allow comparison of different TPC gas and pressure/density choices, we seek to define a {\em tangled track reconstruction figure of merit} $F_{TTR}$.  For a $\theta_t$-tangled track of length $L$ in a detector with spatial resolution $\sigma_x$, measures of the track reconstruction fidelity must have the form:

\[
F_{TTR} \equiv \frac{L(Q)}{\sigma_{x}}\frac{1}{f(\theta_t)} \approx  \frac{L(Q)}{\sigma_x}\frac{1}{1+\theta_t^{1.5}}
\]\label{def_fttr}

where $L(Q)$ is range of a single electron at the \ovbb~endpoint (i.e., like a single-$\beta$ background event in the energy region of interest), $\sigma_{x}$ is the tracking detector's spatial resolution, $\theta_t$ measures the level of track twisting or kinking during scattering off the material, and the function $f(\theta_t)$ represents {\em how sensitive} your tracking is to $\theta_t$; this function will be different for different algorithms; in general it may depend on the behavior of the tail of a reconstruction-error distribution in a way that is hard to assess more generally.  One example of a reconstruction failure mode is the one where a single-beta track has a kink or twist near the {\em beginning} which is misreconstructed as a high-dE/dx Bragg peak representing an end.  In general, it makes sense to require $f(0)=1$, a monotonic rise, and significant effects appearing by $f(2\pi)$ when typical tracks undergo a full circle's worth of bending.  For concreteness, we invent a simple strawman function $f(\theta_t) = 1+\theta_t^{1.5}$; this is meant as a placeholder and is not associated with any particular algorithm.  The impact of $f(\theta_t)$ variants can be seen in Figure \ref{fig_tangling}.

We will call $f(\theta)$ the {\em tangling power}.

  Thus defined, $F_{TTR}$ captures differences between gases in several ways.
  \begin{enumerate}
  \item Track length: A low-stopping-power gas (roughly proportional to molecular mass, with some composition dependence) increases the figure-of-merit since it allows longer tracks and higher $F_{TTR}$.
  \item Bending: If your stopping power comes from an abundance of low-$Z$ atoms, rather than a few high-Z atoms, you have less tangling power and higher $F_{TTR}$.
  \item Endpoint energy: A high-Q double beta decay means longer tracks in the region of interest and higher $F_{TTR}$.
  \end{enumerate}

  $F_{TTR}$ thus informs your detector design choices via the following reasoning.  At fixed gas density $\rho_0$, the figure of merit can be interpreted as an effective spatial resolution.  Suppose that our track-reconstruction algorithm has given us a minimum acceptable effective spatial resolution $F_\mathrm{min}$.  If we find that we have extra resolution (i.e. $F_{TTR} > F_\mathrm{min}$), this gives us permission to increase the TPC gas density to a maximum of $\rho_\mathrm{max} = \rho_0 F_{TTR}/F_\mathrm{min}$, which increases the target atom count.  Thus, for a fixed detector hardware/readout/software system (i.e. fixed $\sigma_x$), and with a fixed $F_\mathrm{min}$ as a performance floor, $F_{TTR}$ is proportional to the maximum obtainable target atom count per unit volume, and thus to the half-life sensitivity in an enriched detector of fixed volume.  The maximum target atom count per volume in a detector using {\em unenriched} material, with isotope abundance $f$, is also a useful figure of merit:

  \[
  F^\mathrm{nat}_{TTR}  \equiv f\times F_{TTR}
  \]

Those figures-of-merit allow you to compare different compounds' ability to measure very long half-lives.  To connect to neutrino physics, we can use our knowledge of \ovbb~physics to ask  ``Which gas allows the highest neutrino mass discovery potential?''.  If $F_{TTR}$ is proportional to the number of target atoms in a fixed volume, under tangled-track-reconstuction conditions, then (for any given neutrino mass) the event rate is proportional to $\Gamma_\mathrm{TTR}$:

\[
\Gamma_\mathrm{TTR}  \equiv F_{TTR} G |\mathcal{M}|^2
\]

where $G$ is the phase space and $\mathcal{M}$ is the nuclear matrix element for \ovbb.   A quantity proportional to the event rate in an unenriched detector is then:

\[
\Gamma^{\mathrm{nat}}_\mathrm{TTR}  \equiv f \Gamma_\mathrm{TTR}
\]

\subsubsection{Figures-of-merits of the new proposed electropositive gas list}

In Table \ref{tab_fom} we show these four figures-of-merit for each of our candidate electropositive gas TPC fluids.  Rather than showing absolute values, we express figures-of-merit as ratios with the pure xenon.  From this table we learn a handful of things: \begin{itemize}
\item  Simpler compounds {\em only slightly} outperform complex ones.  For example, germane outperforms tetramethylgermanium in $F_{TTR}$ by 37\%; this means that, all else being equal, a 72 mol/m$^3$ germane TPC has the same tracking performance as a 53 mol$m^3$ tetramethylgermanium TPC.  (If we had focused on matching track lengths and ignored the tangling effect, 72~mol/m$^3$ germane would be equivalent to only 41~mol/m$^3$ tetramethylgermanium.)  
\item Selenium-based gas TPCs (both enriched and natural) have far higher discovery potential than xenon TPCs, benefitting from longer and straighter tracks (visible in $F_{TTR}$) as well as more favorable \ovbb~physics (visible in $\Gamma_{TTR}$).  
\item Two high-endpoint nuclides, $^{96}$Zr and $^{100}$Mo, occur in large compounds that slightly outperform xenon in tracking despite their much greater stopping powers.  We note that this conclusion is especially sensitive to the $f(\theta_t)$ model.  
\item Germanium TPCs are competitive with xenon, with the tracking benefits of a low stopping power largely cancelling the tracking difficulties of a low endpoint energy.
\item Stannane and tetramethyltin TPCs {\em only slightly} underperform xenon TPCs on all measures.
\item Isotope enrichment will be a huge cost driver for beyond-ton-scale \ovbb.  For xenon, switching from $^{\mathrm{nat}}$Xe to highly-enriched $^{136}$Xe gains a factor of 11.6 in decay rate per unit volume.  This table shows that switching from $^{\mathrm{nat}}$Xe to $^{\mathrm{nat}}$Te also gains a factor of 11 in decay rate per unit volume---probably at much lower cost.  Selenium and molybdenum also have excellent unenriched-detector performance.  
\item If the reader is interested in developing new volatile transition-metal complexes for these uses, we would like to enlarge our library of Mo sandwich compounds.  A new Nd option would be more useful than Ca or Pd.  We beg the reader to ignore Cd whether it is valuable or not.
  \end{itemize}

\begin{table} % table contents generated by python fom.py FOM
      \begin{center}{\bf Figures-of-merit for electron drift gas candidates}\end{center}
      \begin{tabular}{l | r c c |  c c c c | r r }
      \cline{2-10}
 \multicolumn{1}{c}{}     &  \multicolumn{3}{ c|}{Endpoint $\beta$}  & \multicolumn{4}{ c|}{ \ovbb~search}    &   \multicolumn{2}{c}{Operating} \\
  \multicolumn{1}{c}{} &  \multicolumn{3}{ c|}{track features}  &  \multicolumn{4}{ c |}{ figures-of-merit}    & \multicolumn{2}{c}{conditions}  \\
      \cline{2-10}
 \multicolumn{1}{c}{}      & Track & Rad.   &       & $F_{TTR}$ & $F^{\mathrm{nat}}_{TTR}$ & $\Gamma_{TTR}$ & $\Gamma^{\mathrm{nat}}_{TTR}$  & $\rho_\mathrm{op}$ & $p_\mathrm{op}$ \\
  \multicolumn{1}{c}{}         & len.   & len. &     $\theta_t$                & (vs.     & (vs.                  &  (vs.   & (vs.     &  (mol/  & (bar)     \\
  \multicolumn{1}{l}{Compound}    & (cm)     & (cm)   &     &  $^{136}$Xe)& $^{\mathrm{nat}}$Xe)                 & $^{136}$Xe) & $^{\mathrm{nat}}$Xe) &  m$^3$)  &                 \\  
  \hline
Xenon (STP) & 255 & 1467 & 4.6 & 1.00 & 1.00 & 1.00 & 1.00 & 45 & 1.0  \\
\hline
Hydrogen selenide & 493 & 3370 & 3.5 & 2.82 & 2.92 & 8.10 & 8.37 & 126 & >2.8  \\
Carbonyl selenide\textsuperscript{$\star$} & 372 & 3043 & 3.2 & 2.39 & 2.47 & 6.86 & 7.09 & 107 & >2.4  \\
Methane selenol & 417 & 3219 & 3.3 & 2.58 & 2.67 & 7.41 & 7.66 & 115 & >2.6  \\
Selenophene\textsuperscript{\textdagger} & 301 & 2871 & 2.9 & 2.13 & 2.20 & 6.11 & 6.32 & 95 & >2.1  \\
Selenolane\textsuperscript{\textdagger} & 291 & 2847 & 2.9 & 2.10 & 2.17 & 6.01 & 6.21 & 93 & >2.1  \\
Benzeneselenol & 251 & 2668 & 2.8 & 1.90 & 1.97 & 5.45 & 5.63 & 85 & >1.9  \\
1,4-diselenane\textsuperscript{\textdagger$\star$} & 185 & 1547 & 3.1 & 2.41 & 2.49 & 6.91 & 7.15 & 54 & >1.2  \\
Bz 1,4-diselenol\textsuperscript{\textdagger$\star$} & 167 & 1492 & 3.0 & 2.27 & 2.35 & 6.52 & 6.74 & 51 & ?  \\
\hline
Tellurophene\textsuperscript{\textdagger} & 190 & 1447 & 3.9 & 0.93 & 3.62 & 2.92 & 11.31 & 42 & 0.9  \\
Tellurolane\textsuperscript{\textdagger} & 185 & 1442 & 3.9 & 0.92 & 3.58 & 2.88 & 11.18 & 41 & 0.9  \\
\hline
Germane & 351 & 3746 & 4.1 & 1.62 & 1.42 & 1.38 & 1.21 & 72 & >1.6  \\
Tetramethyl Ge & 201 & 3098 & 3.4 & 1.18 & 1.04 & 1.00 & 0.88 & 53 & >1.2  \\
\hline
Stannane\textsuperscript{$\star$} & 251 & 1673 & 4.6 & 0.99 & 0.64 & 1.67 & 1.09 & 44 & 1.0  \\
Tetramethyltin\textsuperscript{\textdagger} & 169 & 1530 & 3.9 & 0.82 & 0.53 & 1.38 & 0.90 & 36 & 0.8  \\
\hline
TDMAZ\textsuperscript{\textdagger} & 163 & 1812 & 2.4 & 1.45 & 0.46 & 3.29 & 1.03 & 65 & >1.5  \\
Zr(O$^t$Bu)$_4$\textsuperscript{\textdagger$\star$} & 113 & 1408 & 2.3 & 1.08 & 0.34 & 2.43 & 0.77 & 48 & >1.1  \\
Zr(OEt)$_4$\textsuperscript{\textdagger$\star$} & 159 & 1670 & 2.5 & 1.37 & 0.43 & 3.09 & 0.97 & 61 & ?  \\
\hline
(EtBz)$_2$Mo\textsuperscript{\textdagger$\star$} & 129 & 1539 & 2.6 & 1.07 & 1.15 & 6.96 & 7.51 & 48 & >1.1  \\
\hline
CaC$_8$H$_{20}$\textsuperscript{\P} & 341 & 4436 & 1.8 & 4.36 & 0.09 & 0.46 & 0.01 & 194 & ?  \\
PdC$_8$H$_{20}$\textsuperscript{\P} & 120 & 1588 & 3.7 & 0.63 & 0.84 & 1.47 & 1.95 & 28 & ?  \\
CdC$_8$H$_{20}$\textsuperscript{\P} & 162 & 1497 & 3.2 & 1.04 & 0.89 & 3.81 & 3.25 & 46 & ?  \\
NdC$_8$H$_{20}$\textsuperscript{\P} & 170 & 1072 & 3.2 & 1.08 & 0.68 & 7.80 & 4.91 & 48 & ?  \\\end{tabular}
      \caption{Key length scales and figures-of-merit for our proposed electropositive gases.  For each gas we give the CSDA electron range (from  ESTAR\cite{ESTAR}) and radiation length (from \cite{tsai1974}) at an STP-like constant 44.6~mol/m$^3$, and the tangling measure $\theta_t$.   These are given for a background-like single electron at the relevant \ovbb~endpoint.  We calculate the four tracking performance figures-of-merit defined in the text, using the  strawman form $f(\theta_t) = 1 + \theta^{3/2}$ of the tangling power.  Each value is given as a ratio with the corresponding one in xenon.  Using the figure-of-merit $F_{TTR}$, we can define a set of operating densities $\rho_\mathrm{op}$ under which the endpoint-energy tracking performance is the same across all the gases listed.  When operating at this tracking-matched $\rho_\mathrm{op}$, $F_{TTR}$ and $F^\mathrm{nat}_{TTR}$ are proportional to the number density of target nuclei; $\Gamma_{TTR}$ and $\Gamma^\mathrm{nat}_{TTR}$ are proportional to the number density of \ovbb~decay rate at a fixed $m_{\beta\beta}$.  Nuclear matrix elements and phase space factors used for $\Gamma$ are from \cite{hyvarinenNuclearMatrixElements2015a, tsunodaShapeTransitionNd2023, simkovicNuclear2013, stoicaPhaseSpaceFactors2019a} variously.  We give the pressure corresponding to $\rho_\mathrm{op}$ assuming operation near the atmospheric boiling point, with the caveat that the vapor pressure curves (and some boiling points) are unknown.  Finally, although we have not identified useful electropositive organometallics for Ca, Pd, Cd, or Nd, we have included stopping calculations for ``generic'' smallish complexes [M]C$_8$H$_{20}$ (labeled with pilcrows\textsuperscript{\P}), to motivate or demotivate future work.}\label{tab_fom}
\end{table}

\begin{figure}
  \begin{center}
    \includegraphics[width=\textwidth]{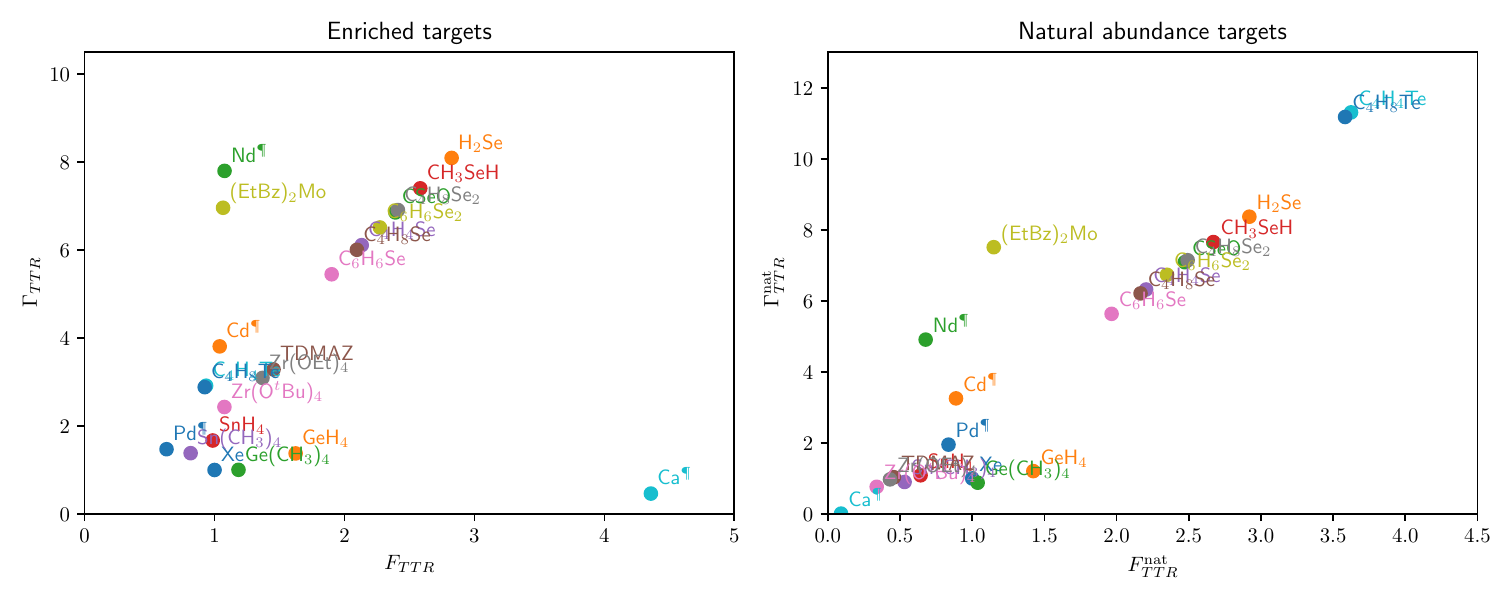}
  \end{center}
  \caption{Scatter plots of the performance figures-of-merit $F_{TTR}$ and $\Gamma_{TTR}$ for the listed compounds, from the numbers in Table~\ref{tab_fom}.  For detectors at densities yielding the same tracking performance, $\Gamma_{TTR}$ is proportional to the \ovbb~event rate density, and $F_{TTR}$ is proportional to the number density of target nuclei.  Different compounds carrying the same nuclide will lie approximately (but not exactly) along a characteristic line ordered by molecular mass.  Versions of this plot under alternate $f(\theta_t)$ models can be found in Figure \ref{fig_FOM_variant} in the appendix.}\label{fig_fom}
  \end{figure}

\subsection{Giant TPCs at 10-, 100-, and 1000-T scale}

Gas TPCs necessarily have larger underground space requirements than dense liquid TPCs.  The size of this footprint depends on the operating pressure/density.  Two novel approaches to obtaining and configuring these spaces have been proposed, with the goal of reducing the construction cost and complexity of using conventional pressure vessels. 

First, if we build detectors that operate at {\em ambient atmospheric pressure}\cite{monrealCPAD2024}, the underground spaces we need grow large but not prohibitively large.  In exchange for this large volume footprint, we hugely simplify the detector mechanics: we can dispense with a pressure vessel entirely.  For a detector using hazardous gases, there are safety benefits to an approach that removes any pressurized gas venting scenarios.  Fortunately, in a TPC whose spatial resolution is diffusion-limited, 1~atm (ambient) pressure appears to be a better choice than high pressure\cite{mistryOptimalOperatingParameters2025,byrnes2024}.  

Second, we have argued\cite{monrealHighpressureTPCsPressurized2022a} that {\em lined rock caverns} (LRCs) would allow large underground high-pressure detectors to be constructed without individual pressure vessels.  Lined rock caverns\cite{lindblom1985} are a gas-storage-industry innovation in which a large underground cavern is given a thin steel skin which allows it to hold high pressure, as much as 250~bar.  When the cavern interior is pressurized, the skin transmits the pressure forces into the rock.  Three real-world LRCs are at Grangesberg (R\&D)\cite{federalenergytechnologycenterpittsburghpausCommercialPotentialNatural1999a}, Skallen (natural gas)\cite{glamheden2005}, and the HYBRIT green steel pilot plant in Svartöberget (H$_2$)\cite{johansson2018}.   By using LRCs to host physics experiments, we can operate design very large TPCs at very high pressures without the ordinary constraints of pressure vessel engineering.   Similar unconventional pressurization solutions might be found underwater, in solution-mined salt caverns, or in flooded levels and shafts \cite{monrealSubPenningGasMixtures2015a} of a conventional mine.

Finally, conventional pressure vessels remain an option.

It is easy to engineer a detector that operates at near-ambient temperature; this seems to favor the subset of gases which are gases near room temperature (Hydrogen selenide, carbonyl selenide, methaneselenol, germane, and stannane).  Colder detectors-than-ambient detectors raise familiar engineering issues; chilling a few tens of degrees may prove to have few or no advantages.   Hot detectors, required by the other gases and compatible with all of them, would be a novelty that requires further study; among other issues we can expect higher noise in hot front-end electronics.

\subsubsection{Gas detector size and pressure scales}

In this section, we wish to motivate future detector design studies, and show the differences between detector scales, tracking performance, and observables across the different gases proposed. Table~\ref{tab_size} is meant to give the reader intuition for the relationship between plausible detector size scales, pressure scales, and target atom count.

The figures-of-merit in the earlier Table~\ref{tab_fom} can be interpreted as the target molar density and/or volumetric decay rate when {\em tracking performance} is held constant.  Table~\ref{tab_300} holds {\em pressure} constant at 1~atm and shows how the densities, decay rates, and tracking features vary over gases.  This is given in a fixed volume (50,000 m$^3$) so it immediately translates to a large but not unimaginable apparatus.  

\begin{table}
  \begin{center}{\bf Illustration of mass/size/pressure scales for large gas TPCs}\end{center}
  \begin{tabular}{c | c c c c}
    
    \multicolumn{2}{r}{\begin{tikzpicture}[x=2,y=2]

         \draw (0,0)--(0,21.5) arc[start angle=0,end angle=180,x radius=17.5,y radius=17] -- (-35,0) arc[start angle=-180,end angle=0,x radius=17.5,y radius=12];
         
         \draw (-3,-10)--(-3,31.4) arc[start angle=0,end angle=180,x radius=19.65,y radius=6] -- (-42.3,-10)--(-3,-10);
         \draw[dashed] (-3,31.4) -- (-42.3,31.4);

\end{tikzpicture}}
    &
    
\begin{tikzpicture}[x=2,y=2]
   \draw (0,0)--(-18,0);
   \draw plot [smooth] coordinates {(-18,0) (-20,14) (-18,28) (-9,33) (0,28) (2,14) (0,0)};
   \draw (5,2) arc[start angle=-36,radius=10,delta angle=252] -- (5,2);
\end{tikzpicture}
& 

\begin{tikzpicture}[x=2,y=2]
  \draw (-11.5,12) -- (3.5,12) -- (3.5,-3) -- (-11.5,-3) -- (-11.5,12);
  \draw[left=1, above=1,gray] plot coordinates {(-12,0) (-9,3) (-5,4.4) (5,4.4) (9,3) (12,0) (9,-3) (5,-4.4) (-5,-4.4) (-9,-3) (-12,0)};
  \draw plot coordinates {(-12,0) (-9,3) (-5,4.4) (5,4.4) (9,3) (12,0) (9,-3) (5,-4.4) (-5,-4.4) (-9,-3) (-12,0)};

\end{tikzpicture}
&

\begin{tikzpicture}[x=2,y=2]
  \draw (-5.2,0)--(-9.6,0)--(-9.6,7) arc[start angle=0,delta angle=-180,radius=-2.2]--(-5.2,0);

  \draw[left=1, above=1,gray] (0,0)--(0,4.15) arc[start angle=0,end angle=90,x radius=2.0,y radius=0.6] -- (-2.0,4.75)--(-2.0,7) -- (-2.3,7)--(-2.3,4.75) arc[start angle=90,end angle=180,x radius=2.0,y radius=0.6] -- (-4.15,0) arc[start angle=-180,end angle=0,x radius=2.075,y radius=0.6];
  \draw (0,0)--(0,4.15) arc[start angle=0,end angle=90,x radius=2.0,y radius=0.6] -- (-2.0,4.75)--(-2.0,7) -- (-2.3,7)--(-2.3,4.75) arc[start angle=90,end angle=180,x radius=2.0,y radius=0.6] -- (-4.15,0) arc[start angle=-180,end angle=0,x radius=2.075,y radius=0.6];

\end{tikzpicture}
\\

    &    SuperK tank & SNO cavern & SNOLab Cube Hall & Grangesberg LRC \\
    & or Skallen LRC & (30\% LNGS Hall A) & (2x KATRIN) & (2x GERDA cryostat)\\
    & 50,000 m$^3$ & 12,000 m$^3$ & 3,300 m$^3$ & 135 m$^3$ \\
    \hline
    target atoms  & \multicolumn{4}{c}{multiple of STP density required (100\% target isotope)}\\
    \hline    
    10$^{29}$ & 0.08 & 0.3 & 1 & 30 \\
    10$^{30}$ & 0.8 & 3 & 10 & liquid-like \\
    10$^{31}$ & 8  & 30 & 100$^{SC?}$ & -- \\
    \hline
    target atoms   & \multicolumn{4}{c}{multiple of STP density required (9.2\% target isotope)}\\
    \hline
    10$^{29}$ & 0.8 & 3.0 & 12 & liquid-like \\ 
    10$^{30}$ & 8.0 & 30 & 120$^{SC?}$ & -- \\
    10$^{31}$ & 80$^{SC?}$ & liquid-like & --  & -- \\
  \end{tabular}
  \caption{Scaling examples, using familiar containers, to illustrate the pressure requirements to contain large quantities of double-beta decay nuclei in the gas phase.  We show detectors containing a total of 10$^{29}$,10$^{30}$,and 10$^{31}$ target nuclei (roughly corresponding to 10~T, 100~T, and kiloton scales), either in the isotope-enriched case (in which it applies to all species) or a 9.2\%-abundance case, as for natural selenium; the reader may easily rescale to other compounds.  If using 1,4-diselenane or Benzene 1,4-diselenol, the number of target nuclei is twice the given value.  The numbers given are the gas molar density as a multiple of the molar density of an ideal gas at STP (i.e., 44.6 mole/m$^3$); these correspond to different pressures depending on the choice of operating temperature, and at higher densities on the gas equation-of-state.   Most of the compounds of interest are probably supercritical above 50$\times$ STP, which may drive changes in Fano factors, conductivity, and the solubility of radioactive background sources.   Above 300$\times$ STP it seems implausible that there could be benefits to the gas over the liquid.   Values below the STP density could be configured, not as actual underpressures requiring a vacuum vessel, but as ambient-pressure TPCs with balance helium.}  \label{tab_size}

\end{table}

\begin{table}
    \begin{center}{\bf Gas selection and $meV$-scale physics reach in 50,000 m$^3$ instruments at ambient pressure}\end{center}
  \begin{tabular}{l | r r r r | r r |  r r } % table contents generated by: python fom.py cavernX
\multicolumn{1}{c}{} &  \multicolumn{4}{c|}{SuperK-like volume} & \multicolumn{2}{c|}{Scenario (a)} & \multicolumn{2}{c}{Scenario (b)} \\
\multicolumn{1}{c}{} &  \multicolumn{4}{c|}{50k m$^3$ (1 atm @ b.p.)} & \multicolumn{2}{c|}{(enriched)} & \multicolumn{2}{c}{(natural)} \\
\cline{2-9}
 \multicolumn{1}{c}{}  & T & Track & $\theta_t$ & Element & Targets & \ovbb & Target & \ovbb\\
 \multicolumn{1}{c}{}  & (C) & length &    & mass & (atoms & rate & (atoms & rate\\
 \multicolumn{1}{l}{Compound}  &   & (cm) &   & (T) & $\times$10$^{-28}$) & (y$^{-1}$) & $\times10^{-28}$) & (y$^{-1}$)\\
\hline
Xenon (STP) & 0 & 254 & 4.6 & 304 & 134.9 & 1.7 & 12.0 & 0.2 \\
\hline
Hydrogen selenide & -41 & 419 & 3.5 & 216 & 158.7 & 5.9 & 14.6 & 0.5 \\
Carbonyl selenide\textsuperscript{$\star$} & -23 & 340 & 3.2 & 200 & 147.3 & 5.5 & 13.5 & 0.5 \\
Methane selenol & 12 & 435 & 3.3 & 175 & 129.2 & 4.8 & 11.9 & 0.4 \\
Selenophene\textsuperscript{\textdagger} & 110 & 422 & 2.9 & 130 & 96.2 & 3.6 & 8.8 & 0.3 \\
Selenolane\textsuperscript{\textdagger} & 135 & 435 & 2.9 & 122 & 90.3 & 3.3 & 8.3 & 0.3 \\
Benzeneselenol & 183 & 418 & 2.8 & 109 & 80.8 & 3.0 & 7.4 & 0.3 \\
1,4-diselenane\textsuperscript{\textdagger$\star$} & 237 & 344 & 3.1 & 196 & 144.4 & 5.3 & 13.3 & 0.5 \\
Benzene 1,4-diselenol\textsuperscript{\textdagger$\star$} & 250\rlap{$^*$} & 320 & 3.0 & 191 & 140.8 & 5.2 & 13.0 & 0.5 \\
\hline
Tellurophene\textsuperscript{\textdagger} & 148 & 292 & 3.9 & 188 & 87.5 & 3.5 & 30.2 & 1.2 \\
Tellurolane\textsuperscript{\textdagger} & 166 & 297 & 3.9 & 181 & 83.9 & 3.4 & 28.9 & 1.2 \\
\hline
Germane & -88 & 238 & 4.1 & 251 & 199.0 & 2.2 & 15.5 & 0.2 \\
Tetramethylgermanium & 43 & 232 & 3.4 & 147 & 116.5 & 1.3 & 9.1 & 0.1 \\
\hline
Stannane\textsuperscript{$\star$} & -52 & 202 & 4.6 & 342 & 166.6 & 3.6 & 9.7 & 0.2 \\
Tetramethyltin\textsuperscript{\textdagger} & 75 & 215 & 3.9 & 217 & 105.8 & 2.3 & 6.1 & 0.1 \\
\hline
TDMAZ\textsuperscript{\textdagger} & 89 & 216 & 2.4 & 162 & 101.7 & 3.0 & 2.8 & 0.1 \\
Zr(O$^t$Bu)$_4$\textsuperscript{\textdagger$\star$} & 250 & 216 & 2.3 & 112 & 70.4 & 2.1 & 2.0 & 0.1 \\
Zr(OEt)$_4$\textsuperscript{\textdagger$\star$} & 237\rlap{$^*$} & 296 & 2.5 & 115 & 72.2 & 2.1 & 2.0 & 0.1 \\
\hline
(EtBz)$_2$Mo\textsuperscript{\textdagger$\star$} & 250\rlap{$^*$} & 247 & 2.6 & 116 & 70.4 & 5.9 & 6.8 & 0.6 \\
\hline
CaC$_8$H$_{20}$\textsuperscript{\P} & 170\rlap{$^*$} & 553 & 1.8 & 66 & 83.1 & 0.1 & 0.2 & 0.0 \\
PdC$_8$H$_{20}$\textsuperscript{\P} & 170\rlap{$^*$} & 194 & 3.7 & 151 & 83.1 & 2.5 & 9.8 & 0.3 \\
CdC$_8$H$_{20}$\textsuperscript{\P} & 170\rlap{$^*$} & 263 & 3.2 & 160 & 83.1 & 3.9 & 6.3 & 0.3 \\
NdC$_8$H$_{20}$\textsuperscript{\P} & 170\rlap{$^*$} & 276 & 3.2 & 207 & 83.1 & 7.8 & 4.7 & 0.4 \\
\hline
 \end{tabular}
 \caption{Comparison of gases in a reference detector design: a 50,000~m$^3$ TPC, holding the pressure constant at 1~bar and operating at the boiling point, filled under one of two scenarios: either (a) highly enriched or (b) natural-abundance target material.  Unlike in Table~\ref{tab_fom}, variations in the track length (given for a single background-like electron at the \ovbb~endpoint) and tangling power $\theta_t$ point to slightly different instrumentation challenges.  Variations in the atom count, either due to (a) the temperature differences or (b) a combination of temperature and the isotope abundance differences, give insight into the different physics sensitivities of different gas choices.  Using the same matrix elements used in Table~\ref{tab_fom}, we predict the \ovbb~event rate for $m_{\beta\beta} = 1$~meV, sensitivity to which would substantially cover the normal mass hierarchy.  If backgrounds are controllable, these numbers suggest many of these configurations have valuable discovery potential, some even without isotope enrichment.  Asterisks flag unknown or invented boiling points.}\label{tab_300}
 \end{table}

\subsubsection{Backgrounds}

Backgrounds require further study.  Even large, dense gas detectors have at best modest self-shielding against external gamma rays; at 30x STP density of H$_2$Se, a 3~MeV gamma ray attenuates in about 2.5~m.  Internal radiocontamination is interesting; unlike a liquid-phase detector which can bring metals into solution, most radioactive species cannot be readily volatilized.  However, if made volatile, contaminants may leave tracks in the detector in several ways (beta, conversion electron, and gamma) with topologies which may be difficult or impossible to distinguish from signal events.  In particular, $\gamma$ and $\beta/\gamma$ cascades can mimic the topology of a two-track \ovbb~event if one or two gammas eject conversion electrons.  In the absence of proven ultrapurification techniques for these new gases, we may have to tolerate substantial chemical contamination from closely-related volatile species; we restrict our attention to this issue.  In Table \ref{tab_backgrounds} we survey our species-of-interest and their most likely contaminants in a first attempt to identify any ``showstopper'' species at the endpoint.  Of particular concern are the removal of organolead contamination from tin and the removal of $^{82}$Se from tellurium.

\begin{table}
  \begin{tabular}{|m{0.13\textwidth} | m{0.18\textwidth} | m{0.25\textwidth} | m{0.3\textwidth}|}
    \hline
    Target compound & Contaminants & High-energy radioactivities & Mitigation\\
    \hline
    Tin & tetramethyl lead & From $^{210}$Pb, $\beta/\gamma$ up to 1532 keV & Mostly off-endpoint unless combinatoric; clean feedstocks, gas distillation\\
    \hline
    Zirconium  & organohafnium & Cosmogenic $^{178\mathrm{m}}$Hf (31d, 2.45 MeV) & Probably tolerable off-endpoint background\\
    \hline
    Selenium  & none notable & $^{82}$Sr(n,$\gamma$)$^{83}\mathrm{Se}$ (22~m, 3.67 MeV $\beta/\gamma$ decay) & May have irreducible topologies\\
    \hline
    Tellurium & organoselenium & $^{82}$Se \tvbb is irreducible background to $^{130}$Te \ovbb & Clean feedstock, gas distillation\\
    \hline
    Germanes & none notable & & Feedstock ultrapurificaton adequate\\
    \hline
    Molybdenum & Cr, W diarenes & Cosmogenic $^{53}$Cr(n,pn)$^{52}$V and $^{54}$Cr(n,pn)$^{53}$V & Clean feedstocks, gas distillation, materials screening\\
    \hline
   \multirow{10}*{All} &  \multicolumn{2}{c|}{Radon and radon daughters} & Daughter ion transport; materials screening; gamma line vetos? \\
    \cline{2-4}
&  $^{220}$Rn$\rightarrow^{208}$Tl & 4.999 MeV $\beta/\gamma$, 3m & Delayed coincidence with parent\\
    \cline{2-4}
  & $^{222}$Rn$\rightarrow^{214}$Bi  & 3.269 MeV $\beta/\gamma$, 19m &  Prompt alpha coincidence with $^{214}$Po (163$\mu$s)\cite{lahaieStudyRadonBackground2025}\\
    \cline{2-4}
 &  $^{222}$Rn$\overset{\text{rare}}{\rightarrow}$ $^{210}$Tl & 5.482 MeV $\beta/\gamma$, 1.3m  & Delayed coincidence with parent\\
    \cline{2-4}
    & \multicolumn{2}{c|}{Other volatiles}  & Materials screening; use lowest tolerable operating temperature \\
    \hline
  \end{tabular}
  \caption{Background issues raised by likely chemical impurities in the proposed TPC gases, pointing to particular congener-selective purification challenges for tin and tellurium TPCs.  Some contaminants have problematic short-lived cosmogenic daughters which may set an upper limit on the permissible level of the contaminant.  We have not surveyed for (n,$\gamma$) backgrounds.  As always, radon and its daughters require careful study.}\label{tab_backgrounds}
\end{table}

\subsubsection{Safety issues}
    
Finally, detector engineering will be complicated by the fact that these gases are toxic and flammable\cite{avasthiCPAD2024}.  Although the issues are largely (but not entirely, c.f.\cite{gerkeSuppressionElectricalBreakdown2022,caoNnDEx100ConceptualDesign2023}) new to physicists, chemical-industry standards exist which should allow safe design of underground safety systems, including: capacity for clean off-detector storage of the fluid; capacity to buffer any detector fluid pressure/volume changes; capacity to vent unplanned releases, including catastrophic ones, into a flue-scrubbing system that prevents environmental contamination; for an enriched detector, capacity to recover valuable isotope products from flue-scrubbing systems.  On a smaller scale, we need to develop infrastructure and practices for safe benchtop R\&D\cite{aneeshaavasthiPhDThesisPreparation2026}.

\section{Conclusion}

By scanning the chemistry literature and performing some new DFT calculations, we identified 18 chemical compounds, most never before considered, which may allow \ovbb~target nuclides to be carried in an electropositive gas; the candidates cover the nuclides $^{76}$Ge, $^{82}$Se, $^{96}$Zr, $^{100}$Mo, $^{124}$Sn, and $^{130}$Te.  Operating these gases as single-phase gas TPCs, at ambient to high pressures, is a new route to beyond-ton-scale, even kiloton scale, \ovbb~searches in multiple isotopes.  Unlike the xenon case, they do not rely on expensive or globally scarce materials.  Under the assumption that a large gas TPC relies heavily on track topology for background rejection, we devise a ``tangled track reconstruction'' figure of merit which allows us to compare the uses of these gases.  All are seen to outperform $^{136}$Xe in some way.  In particular, several selenium compounds (led by H$_2$Se, carbonyl selenide, and methane selenol) may offer powerful high-endpoint searches, with bis(ethylbenzene)molybdenum not far behind.   Among options for un-enriched detector materials, tellurophene may offer the highest discovery potential.  

\section{Acknowledgements}

We are grateful to Shane Parker for help getting started with DFT calculations in ORCA; to Thomas Gray for general organometallic chemistry advice; to Marc Rubin, Madeline Crowley, and Richard Bihary for chemical safety help; and to Nicholas Byrnes, Ben J.~P.~Jones,  Dinesh Loomba, Dave Nygren, and Sven Vahsen for helpful discussions of TPC physics; and two anonymous referees whose feedback strengthened the paper.  The NEXUS code is created and maintaned by the NEXT collaboration and used with permission.  Igor Novak kindly shared ORCA input files for one compound.  This work was funded in part by the Department of Energy under award DE-SC0020433.  Computing resources were provided by the CWRU High Performance Computing Center.  The authors' research, coding, writing, copyediting, and spellchecking were entirely human effort with no AI assistance.

\bibliography{ovbb_gases.bib}
\bibliographystyle{JHEP}

\section*{Appendix}

\begin{figure}[h]
  \begin{center}
    \includegraphics[width=\textwidth]{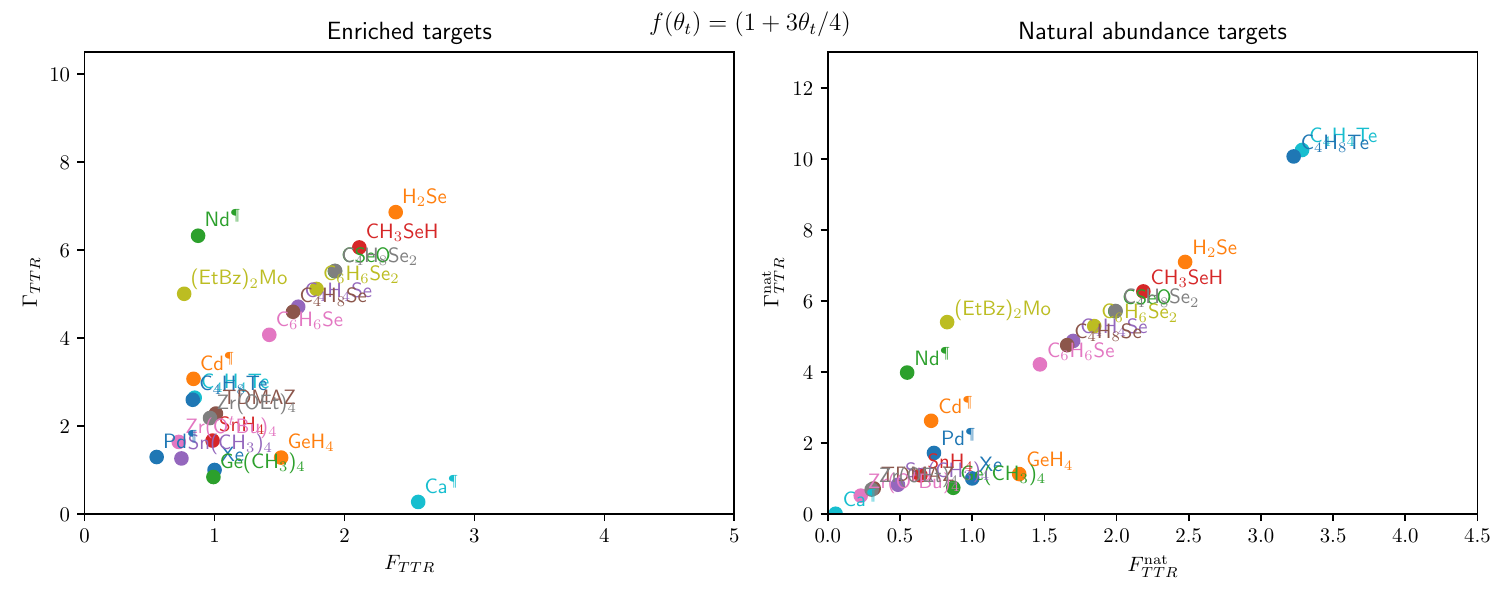}
    \includegraphics[width=\textwidth]{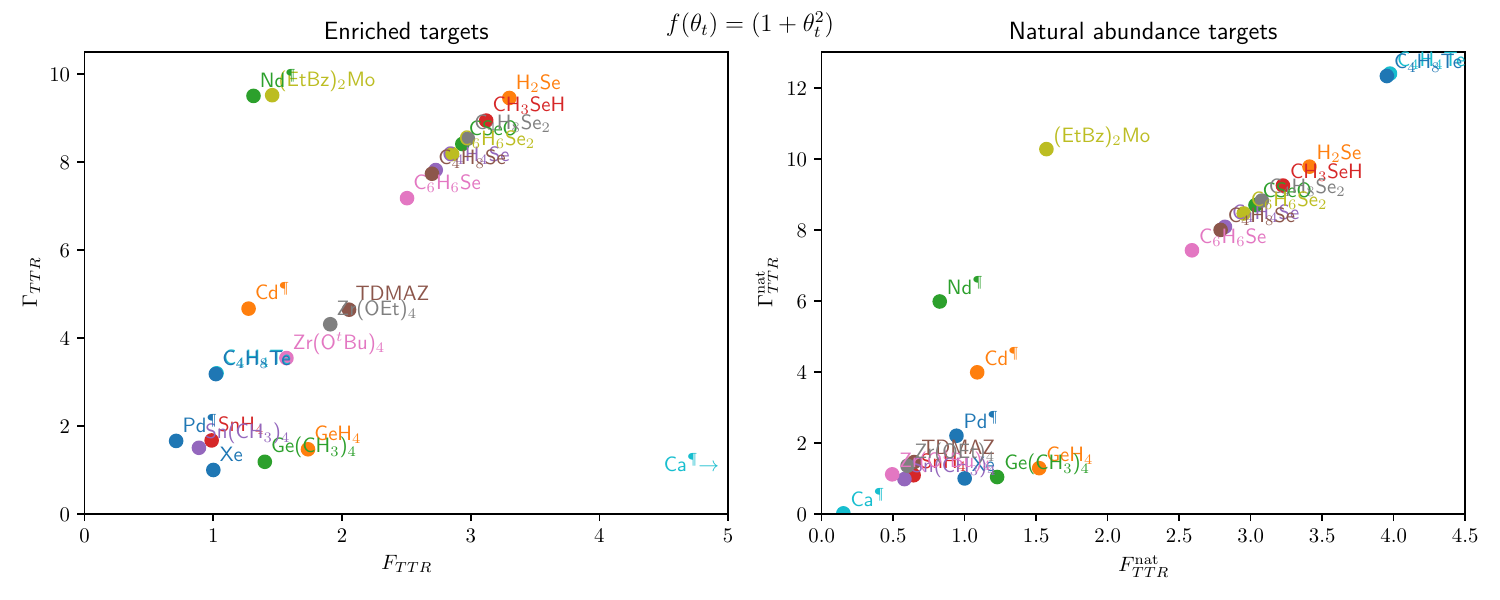}
    \end{center}
  \caption{Figures-of-merit based on two alternate versions of the tangling power formula.  Top panels, $f(\theta_t) = (1 + 2\theta_t)$. Bottom panels,  $f(\theta_t) = (1 + \theta_t^2)$.  For further explanation, see the caption of Figure \ref{fig_fom}, which gives these figures-of-merit using the strawman $f(\theta_t)$ used elsewhere in the paper.}\label{fig_FOM_variant}
  \end{figure}

\begin{table}[h]
  \begin{center}{\bf Electron affinity: DFT comparisons to previous results}
  \begin{tabular}{l| r|r|p{0.25\textwidth} }
    & our value  & Reported & Comments\\
   Reaction           &  (eV) & value (eV) & \\
   \hline   
   O$_2$ $\rightarrow$ O$_2^-$ & +0.86 & +0.45 & \cite{ervinOnlyStableState2003} \\
  CO$_2$ $\rightarrow$ CO$_2^-$ & -0.76 & -0.66 & \cite{gutsevElectronAffinitiesCO21998a} Proxy for CSe$_2$.\\
   CS$_2$ $\rightarrow$ CS$_2^-$ & +0.18 & +0.3 & \cite{gutsevElectronAffinitiesCO21998a} Proxy for CSe$_2$.\\
   CH$_3$NO$_2$  $\rightarrow$  CH$_3$NO$_2^-$ & -0.18 & +0.172 & \cite{adamsLowenergyPhotoelectronImaging2009}\\
   HSe$_2$  $\rightarrow$ HSe$^-$ + H & -1.38 & -1.2 & \cite{abouafLowEnergyElectron2008} Experimental fragment appearance energy; directly relevant to H$_2$Se\\
   SCH$_2$ $\rightarrow$  SCH$_2^-$ & +0.496 & +0.465 & \cite{moranElectronSpectroscopyCH2S1987}\\
   SeF$_6$ $\rightarrow$ SeF$_6^-$ & +3.32 & +2.9 & \cite{comptonCollisionalIonizationFast1978}\\
    CH$_3$SSCH$_3$ $\rightarrow$ CH$_3$SSCH$_3^-$ & +0.24 & +0.23 & \cite{GasphaseIonmoleculeReactions}\\
  C$_6$H$_5$SH $\rightarrow$ C$_6$H$_5$S$^-$ + H & -1.38 & -0.7 & \cite{spyrouElectronswarmMassspectrometricStudy1999} Experiment reports mean electron energy of attaching swarm. Proxy for selenophene.\\
    \hline
  \end{tabular}
  \end{center}
  \caption{While performing DFT calculations for the electron affinity (EA) or dissociative attachment (DEA) threshhold of our compounds of interest, we also modeled similar processes in congener molecules, some of which have experimental or theoretical EA information in the literature.  Those results are tabulated here for comparison.  Negative values mean the reaction requires energy input.  The results suggest that 0.3--0.5~eV uncertainty should be assumed for our other results.}\label{tab_EA}
  \end{table}
\end{document}